\documentclass{article}
\usepackage{graphicx}

\usepackage{amsmath,amssymb,latexsym}

\newcommand{\be}{\begin{equation}}
\newcommand{\ee}{\end{equation}}
\renewcommand{\vec}[1]{\mathbf{#1}}
\begin{document}

\title{Representation of signals as series of orthogonal functions} 
\author{E Aristidi}
\date{\small\sl Laboratoire Lagrange, Universit\'e C\^ote d'Azur, Observatoire de la C\^ote d'Azur, CNRS, Parc Valrose, 06108 Nice Cedex 2, France -- email: \tt aristidi@unice.fr}
\maketitle
\section*{Abstract}
This paper gives an introduction to the theory of orthogonal projection of functions or signals. Several kinds of decomposition are explored: Fourier, Fourier-Legendre, Fourier-Bessel series for 1D signals, and Spherical Harmonic series for 2D signals. We show how physical conditions and/or geometry can guide the choice of the base of functions for the decomposition. The paper is illustrated with several numerical examples.
\section{Introduction}
Fourier analysis is one of the most important and widely used techniques of signal processing. Decomposing a signal as sum of sinusoids, probing the energy contained by a given Fourier coefficient at a given frequency, performing operations such as linear filtering in the Fourier plane: all these techniques are in the toolbox of any engineer or scientist dealing with signals.

However the Fourier series is only a particular case of decomposition, and there exist an infinity of Fourier-like expansions over families of functions. The scope of this course is to make an introduction to mathematical concepts underlying these ideas. We shall talk about Hilbert spaces, scalar (or inner) products, orthogonal functions and orthogonal basis. We shall also see that some kind of signals are well-suited to Fourier decomposition, others will be more efficiently represented as series of Legendre polynomials or Bessel functions. Note that this course is not a rigorous mathematical description of the theory of orthogonal decomposition. Readers who whish a more academic presentation may refer to textbooks in the reference list.

The paper is organized as follows. Basics of Fourier series are introduced in Sect.~\ref{par:fourier}. A parallel is made between Fourier series and decomposition of a vector as linear combination of base vectors. In Sect.~\ref{par:legendre} we generalise the concept to Fourier-Legendre series and present Legendre polynomials as a particular case of the family of orthogonal polynomials. Sect.~\ref{par:SH} is devoted to the expansion of two-dimensional functions of angular spherical coordinates as series of Spherical Harmonics. Finally Sect.~\ref{par:bessel} presents another example of decomposition using sets of Bessel functions.

%
%
\section{Fourier series}
\label{par:fourier}
First ideas about series expansions arise at the beginning of the 19th century. Pionneering work by Joseph Fourier about the heat propagation (Fourier, \cite{fourier}) played a fundamental role in the development of mathematical analysis. Heat propagation is described by a second-order partial differential equation (PDE). To integrate this equation, Fourier proposed to represent solutions as trigonometric series (denominated today as ``Fourier series''). He laid the foundations of the so-called Fourier analysis, which is now extensively used in a wide range of physical and mathematical topics. A modern and comprehensive presentation of Fourier series can be found in Tolstov (\cite{tolstov}).
\subsection{Fourier expansion of a periodic signal}
The basic idea is that a periodic signal $f(t)$ can be approached by a sum of trigonometric functions. The complex form of the expansion is
\be
f(t)=\sum_{n=-\infty}^\infty c_n \; \exp\left(\frac{2i\pi n t}{T}\right)
\label{eq:seriefourier}
\ee
where $T$ is the period of the signal. Discussions about the validity of the development can be found in Tolstov (\cite{tolstov}).  The functions $\phi_n(t)=\exp\left(\frac{2i\pi n t}{T}\right)$ are the {\em harmonic} components of frequency $\frac n T$ present in the signal, and the complex coefficient $c_n$ is a weight. This relation suggests that the ensemble of the coefficients $\{c_n\}$ and the period $T$ contain the same information than the signal itself. High frequency harmonics (large $n$) are generally associated to short-scale variations of the function $f$ (i.e. small details).

Values of the coefficients $c_n$ may be found using the following integral formula (indeed a relation of {\em orthogonality} of the harmonics, as it will be discussed later)
\be
\int_0^T \frac 1 T \: \phi_n(t)\: \overline{\phi_m(t)}\: dt \; = \; \delta_{mn}
\label{eq:orthfourier}
\ee
with $\delta_{mn}$ the Kronecker delta. The notation $\bar \phi$ means the complex conjugate of $\phi$. Combining this formula with Eq.~\ref{eq:seriefourier} gives
\be
c_n\; = \; \int_0^T \frac 1 T \: f(t)\: \overline{\phi_n(t)}\: dt
\label{eq:coefcn}
\ee
\subsection{Example}
\begin{figure}
\includegraphics[width=65mm]{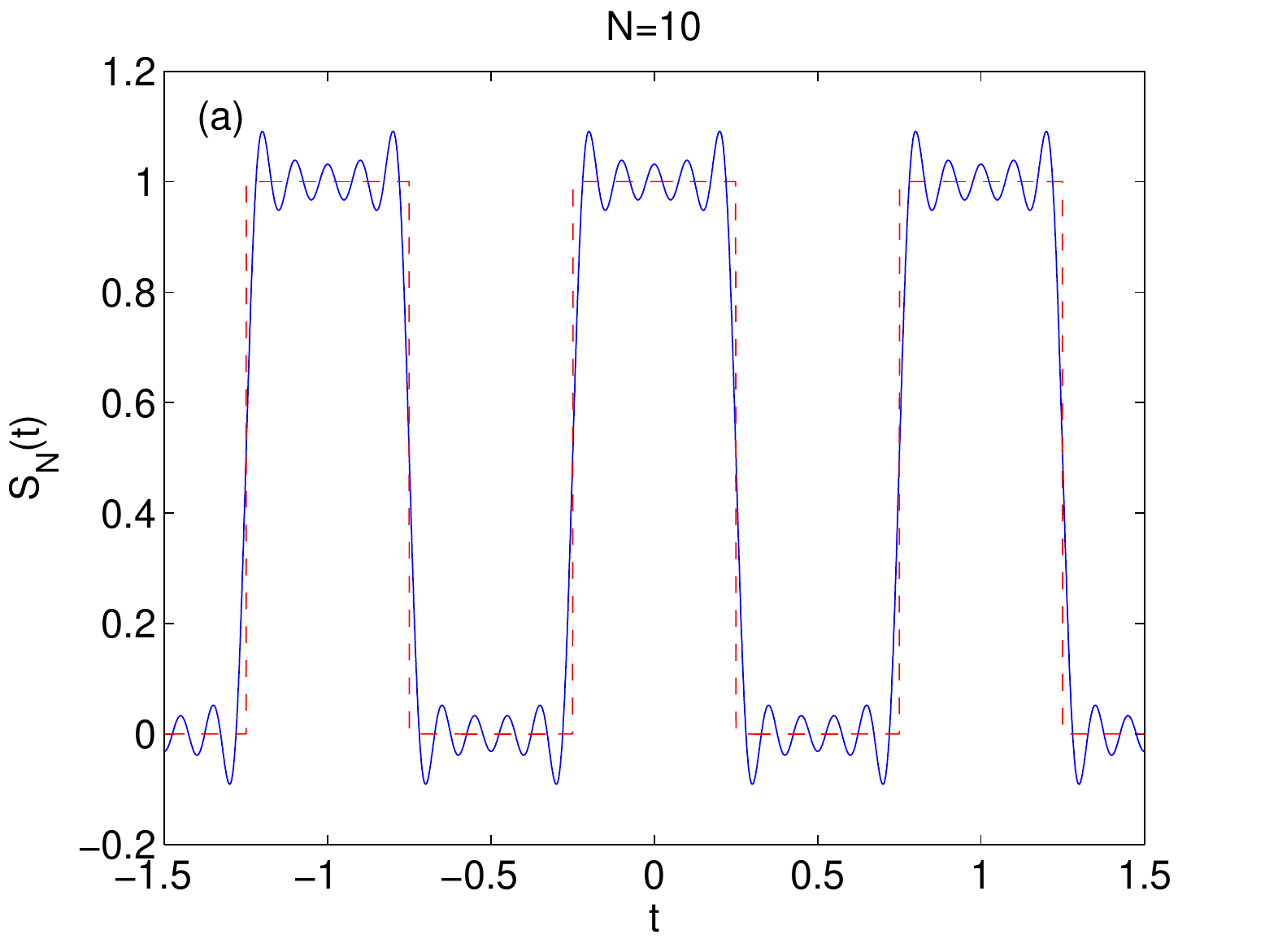}\ \includegraphics[width=65mm]{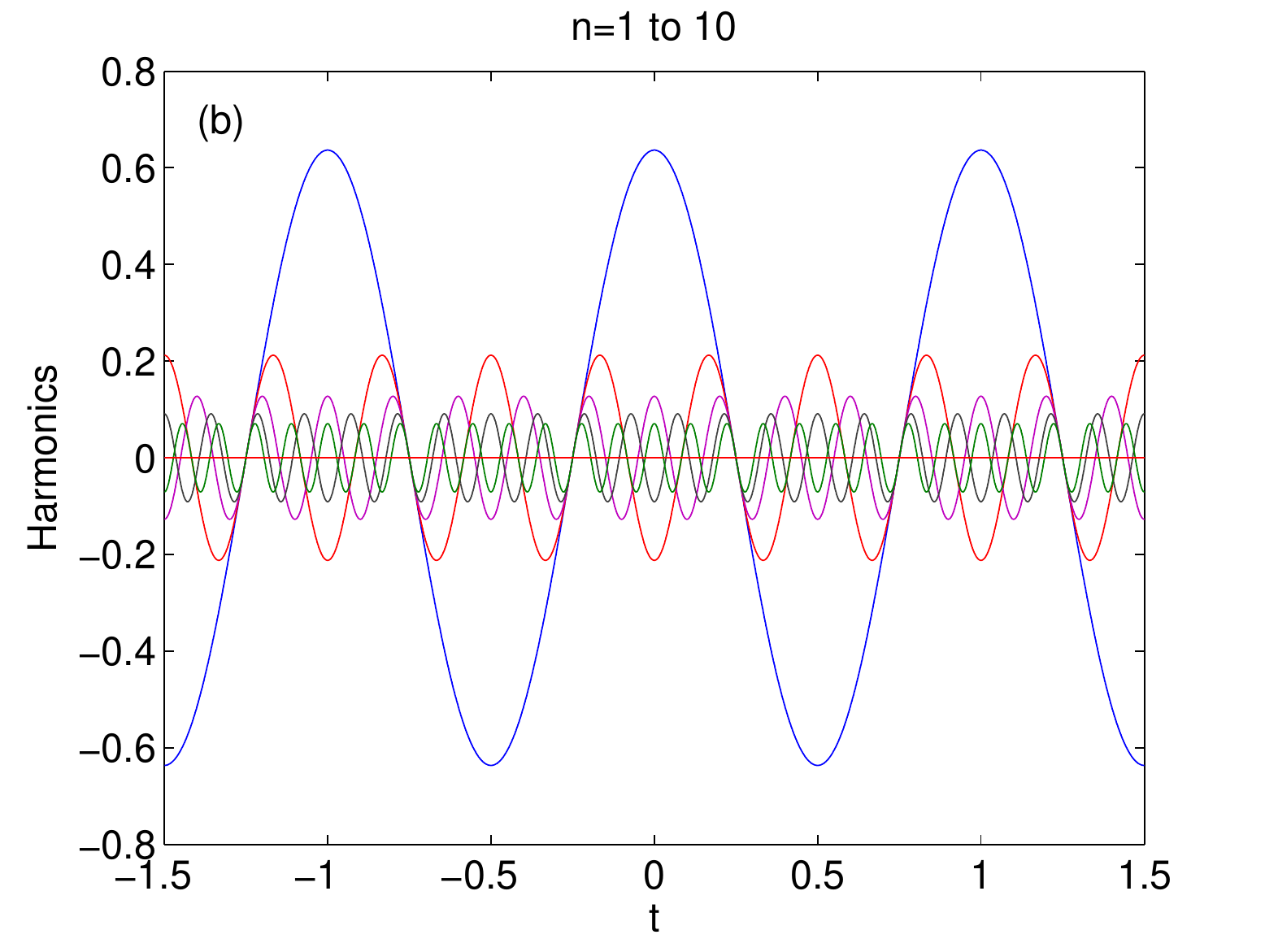}
\includegraphics[width=65mm]{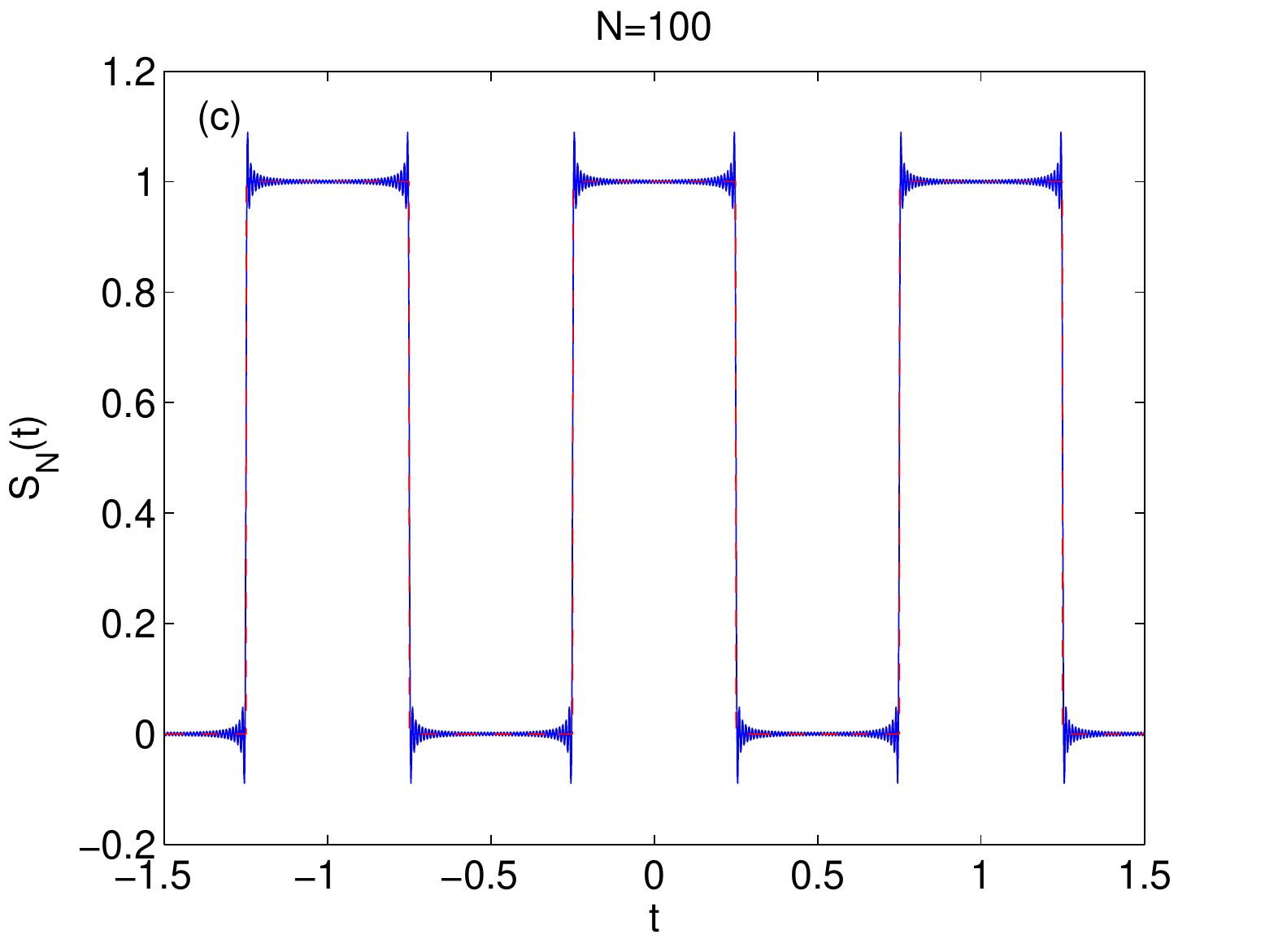}\ \includegraphics[width=65mm]{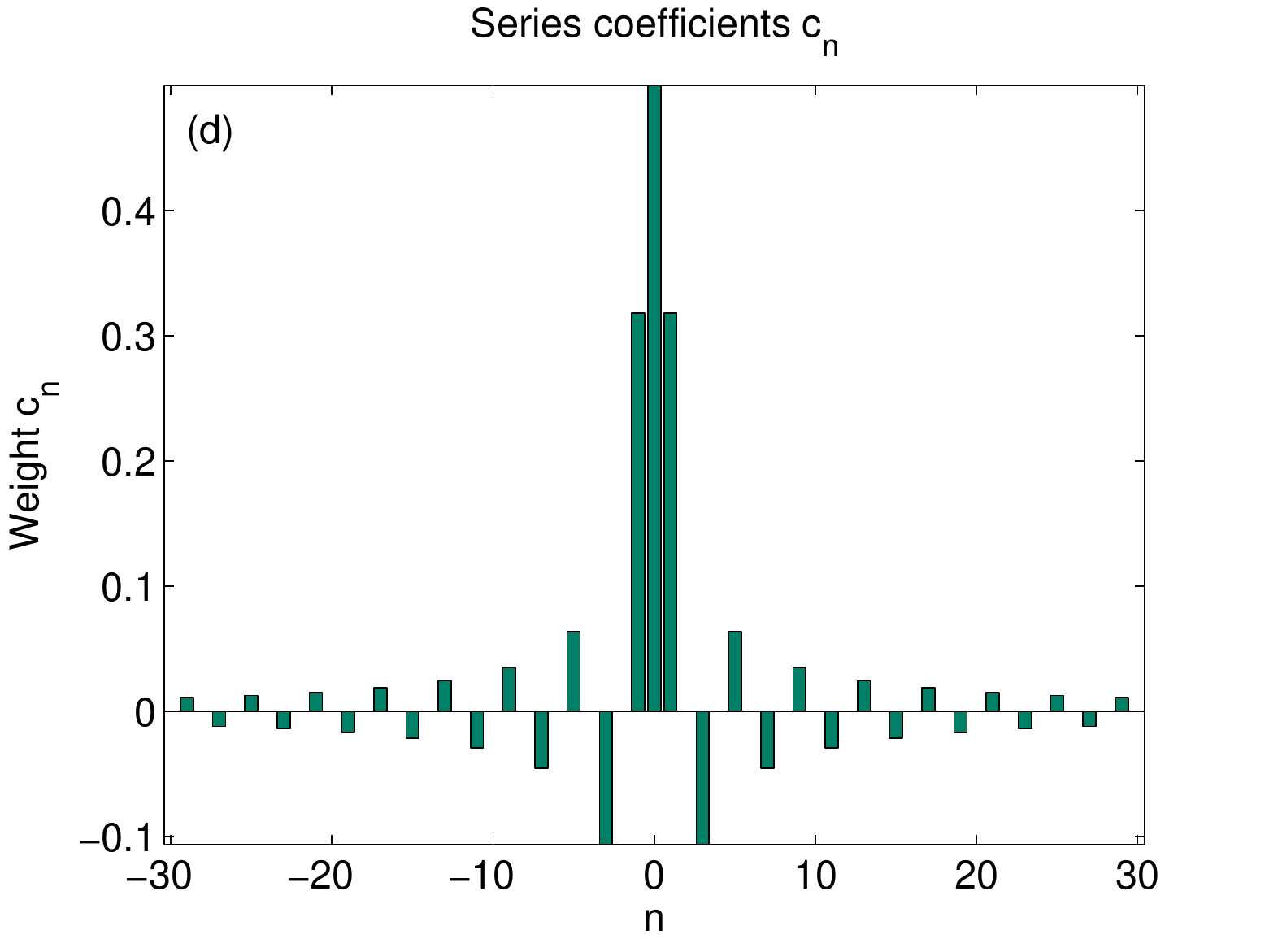}
\caption{Fourier series of a square wave of period $T=1$. (a) Signal (red dashed line) and partial Fourier series with $N=10$ (blue  line). (b) Real part of the terms $c_n \exp^{2i\pi n t}$ (weighted harmonics) for $n=1$ to 10. (c) Partial Fourier series with $N=100$. (d) Graph of $c_n$ versus $n$ (all even terms vanish except $n=0$).}
\label{fig:creneau}
\end{figure}

We consider a square wave of period $T=1$ having value $f(t)=1$ if $|t|<\frac 14$ and 0 elsewhere in the interval $[-\frac12, \frac12]$ (Fig.~\ref{fig:creneau}a). The Fourier coefficients $c_n$ calculated from Eq.~\ref{eq:coefcn} are
\be
c_n=\frac 12 \: \mbox{sinc}\left(\frac{n}{2}\right)
\ee
with sinc$(x)=\frac{\sin(\pi x)}{\pi x}$ the normalized sinc function. The signal $f(t)$ can thus be approached by the sum
\be
S_N(t)=\sum_{n=-N}^N \frac 12 \: \mbox{sinc}\left(\frac{n}{2}\right) \; e^{2i\pi n t}
\ee
$S_N(t)$ is a partial Fourier series which tends towards $f(t)$ as $N\rightarrow\infty$. Fig~\ref{fig:creneau}a and \ref{fig:creneau}b show, for $N=1$ to 10, the real part of the term $c_n e^{2i\pi n t}$ and the partial sum $S_N(t)$. For $N=100$ (Fig~\ref{fig:creneau}c), the convergence seems fairly good, excepted at the points of discontinuity where it presents oscillations (known as the {\em Gibbs phenomenom}). These oscillations will vanish as $N$ increases for any point, except the discontinuity itself. Analysis of the Gibbs phenomenom was given by B\^ocher (\cite{bocher}).

A plot of Fourier coefficients $c_n$ versus $n$ is displayed in Fig~\ref{fig:creneau}d. As it was said, the information contained in this graph and in the signal are the same ($c_n$ is real here, and has no imaginary part). For this example it can be seen that $c_n$ decreases at large $n$, which is quite commun for signals in physics. The decay is low because of discontinuities, but would be faster for smooth signals.

\subsection{Fourier series as a scalar product}
\label{par:scalarprod}
Representation of functions as sum of harmonics has strong analogies with the decomposition of a vector over an orthogonal basis. An historical introduction of these ideas can be found in Birkhoff \& Kreysgig (\cite{birkhoff}) and references therein. In this section we will pick up some analogies between Fourier series and representations of vectors in the 3-dimensionnal space $\mathbb{C}^3$ (vectors with complex components).

\subsubsection{Scalar product and norm}
\paragraph{Vectors}
Consider two vectors $\vec{u}=(u_1, u_2, u_3)$ and $\vec{v}=(v_1, v_2, v_3)$ in $\mathbb{C}^3$ where  $u_k$ and $v_k$ are complex numbers. Their scalar product (indeed {\em hermitian scalar product} for complex vectors) is the complex number
\be
\vec{u}. {\vec{v}}=\sum_{i=1}^3 u_i \bar{v_i}
\ee
The norm of the vector $\vec{u}$ is
\be
\Vert \vec{u}\Vert^2=\vec{u}.{\vec{u}}
\ee
Minimum conditions to fulfill for a (hermitian) scalar product is to be 
\begin{itemize}
\item bilinear, in the sense of  $(\lambda \vec{u}).\vec{v}=\lambda (\vec{u}.\vec{v})$, and  $ \vec{u}.(\lambda\vec{v})=\bar\lambda (\vec{u}.\vec{v})$ (linear in $u$, antilinear in $v$). $\lambda$ is a constant.
\item conjugate symmetric: $\vec{u}. \vec{v}= \overline{\vec{v}. \vec{u}}$ 
\item definite positive: the norm of a vector is always positive. If it is zero, then the vector itself is zero. 
\end{itemize}

\paragraph{Functions}
For two periodic functions $f$ and $g$ with period $T$, the following integral
\be
\langle f,g\rangle \: = \: \int_0^T \frac 1 T \: f(t)\: \overline{g(t)}\: dt
\label{eq:scalarprodfou}
\ee
is has the 3 properties defined above. It is the generalization for functions of the concept of scalar product, it is generally denoted as the inner product of the functions $f$ and $g$. These functions are treated as vectors in a space of functions (Hilbert space).

The associated norm is 
\be
\langle f,f\rangle \: = \: \int_0^T \frac 1 T \: |f(t)|^2\: dt
\ee
This definition for the scalar (inner) product is well suited to Fourier series.  As we will see later, the definition is likely to change depending on the kind of decomposition, in particular the weigthing term $\frac 1 T$ inside the integral.
\subsubsection{Orthogonality}
Two vectors $\vec{u}$ and $\vec{v}$ are orthogonal if their scalar product $\vec u.\vec v$  is zero. Similarly, two functions $f$ ang $g$ are said to be orthogonal if their scalar (inner) product $\langle f,g\rangle \: = 0$.
\subsubsection{Orthonormal base}
\paragraph{Vectors}
For 3 dimensional vectors of $\mathbb{C}^3$, an orthonormal base is composed of 3 unit vectors ($\hat x, \hat y, \hat z$) verifying
\be
\left[
\begin{array}{l}
\hat x.\hat y=\hat x.\hat z=\hat y.\hat z=0\\ 
\hat x.\hat x=\hat y.\hat y=\hat z.\hat z=1
\end{array}
\right.
\ee
The number of {\em base vectors} required to construct an complete orthonormal base is the dimension of the vector space, here 3.

\paragraph{Functions}
This definition can be extended to periodic functions. Eq.~\ref{eq:orthfourier} is indeed the scalar product of the harmonics $\phi_n$ and $\phi_m$ and may be rewriten as
\be
\langle \phi_n,\phi_m\rangle \: =\; \delta_{mn}
\ee
and suggests that the harmonics $\phi_n(t)=\exp\left(\frac{2i\pi n t}{T}\right)$ form an orthonormal base of the space of periodic functions of period $T$. Like the number of base functions, the dimension of this space in infinite. The number of base functions is infinite, the base is said to be complete if every function of this space can be written as a linear combination on $\phi_n$.

\subsubsection{Orthogonal decomposition}
\paragraph{Vectors}
Any vector $\vec{u}$ of $\mathbb{C}^3$ can be written as a weighted sum of the base vectors, 
\be
\vec{u}=u_1\, \hat x\: +\: u_2\, \hat y\: +\: u_3\, \hat z
\ee
where the numbers $u_k$ are given by the scalar products
\be
u_1=\vec{u}.\hat x \hskip 1cm u_2=\vec{u}.\hat y \hskip 1cm  u_3=\vec{u}.\hat z
\ee
i.e. the {\em projection} of $\vec{u}$ on the base vectors. 
\paragraph{Functions}
Analogy to periodic functions is straightforward : the definition of the Fourier series of Eq.~\ref{eq:seriefourier} is indeed a decomposition of the function $f$ on the base vectors $\phi_n$:
\be
f(t)=\sum_{n=-\infty}^\infty c_n \: \phi_n(t)
\ee
where $c_n$ is given by the integral relation of Eq.~\ref{eq:coefcn} which is exactly the scalar product between $f$ and the base function $\phi_n$.

\subsection{Fourier series and differential equations}
\label{par:fourierserDE}
Fourier series were introduced to solve the differential equation (DE) of heat propagation. It is a PDE whose one dimensionnal form is
\be
\frac{\partial u}{\partial t}-\alpha \frac{\partial^2 u}{\partial x^2}=0
\label{eq:diffusion}
\ee
The solution $u(x,t)$ is a function of the time $t$ and a space coordinate $x$. $\alpha$ is a positive real number. A classical technique to solve this kind of equation is to look for solutions of the form $u(x,t)=X(x).T(t)$ where the dependence on $x$ and $t$ is separated. Comprehensive presentations of this method of separation of variables can be found in Mathews and Walker (\cite{walker}) and in section~2 of Jackson (\cite{jackson}). Substituting $u$ back into equation~\ref{eq:diffusion} one finds
\be
\frac{X''(x)}{X(x)}=\frac{T'(t)}{\alpha T(t)}
\ee
Since both sides of the equation depend on different variables, they must then be equal to some constant $-\lambda$. The equation for $X$ becomes
\be
X''+\lambda X=0
\label{eq:diffusX}
\ee
If $\lambda$ is positive then Eq.~\ref{eq:diffusX} is the harmonic equation. Base solutions are $X(x)=e^{\pm i k x}$ where $k=\sqrt{\lambda}$. $k$ is a constant generally determined by boundary conditions. It takes often the form $k=2\pi\frac n a$ where $n$ is an integer and $a$ a characteristical length (if one studies the heat propagation inside a bar, then $a$ is the length of the bar). Hence the general solution of Eq.~\ref{eq:diffusX} is the linear combination of the base solutions for any $n$:
\be
X(x)=\sum_{n=-\infty}^\infty c_n\: e^{2 i \pi n \frac x a}
\ee
which is exactly a Fourier expansion of the function $X(x)$.

%
%
\section{Legendre polynomials}
\label{par:legendre}
\subsection{Introduction}
\label{par:introlegendre}
Legendre polynomials appear for the first time in the work of Legendre~(\cite{legendre}) in relation to problems of celestial mechanics. He proposed series expansion of the Newtonian potential. 
Consider a punctual mass at position $\vec{r}'$, the potential at position $\vec{R}$ is
\be
\Phi(\vec R)\propto \frac{1}{|\vec{R}-\vec{r}'|}\; = \; \frac 1 R \left[1 -2 r\cos\theta +r^2 \right]^{-\frac1 2}
\ee
where $\theta$ is the angle between the vectors $\vec{R}$ and $\vec{r}'$ and $r=\frac{r'}{R}$. A Taylor expansion in $r$, valid for $r<1$ , is
\be
\left[1 -2 r\cos\theta +r^2 \right]^{-\frac1 2} \; = \; \sum_{n=0}^\infty r^n\: P_n(\cos\theta)
\label{eq:devlegendre}
\ee
The coefficients $P_n(\cos\theta)$ are polynoms of degree $n$, known today as Legendre polynomials. Eq.~\ref{eq:devlegendre} allow for example to write the potential created by any mass (or charge) distribution as a series solution involving integrals of the mass (or charge) density function (multipole expansion). See Jackson,~(\cite{jackson}) for a more complete presentation.
The function 
\be
g(x,r)=\left[1 -2 r x +r^2 \right]^{-\frac1 2} 
\ee
is called {\em generating function} for Legendre polynomials.

\subsection{Legendre polynomials}
\begin{figure}
\includegraphics[width=65mm]{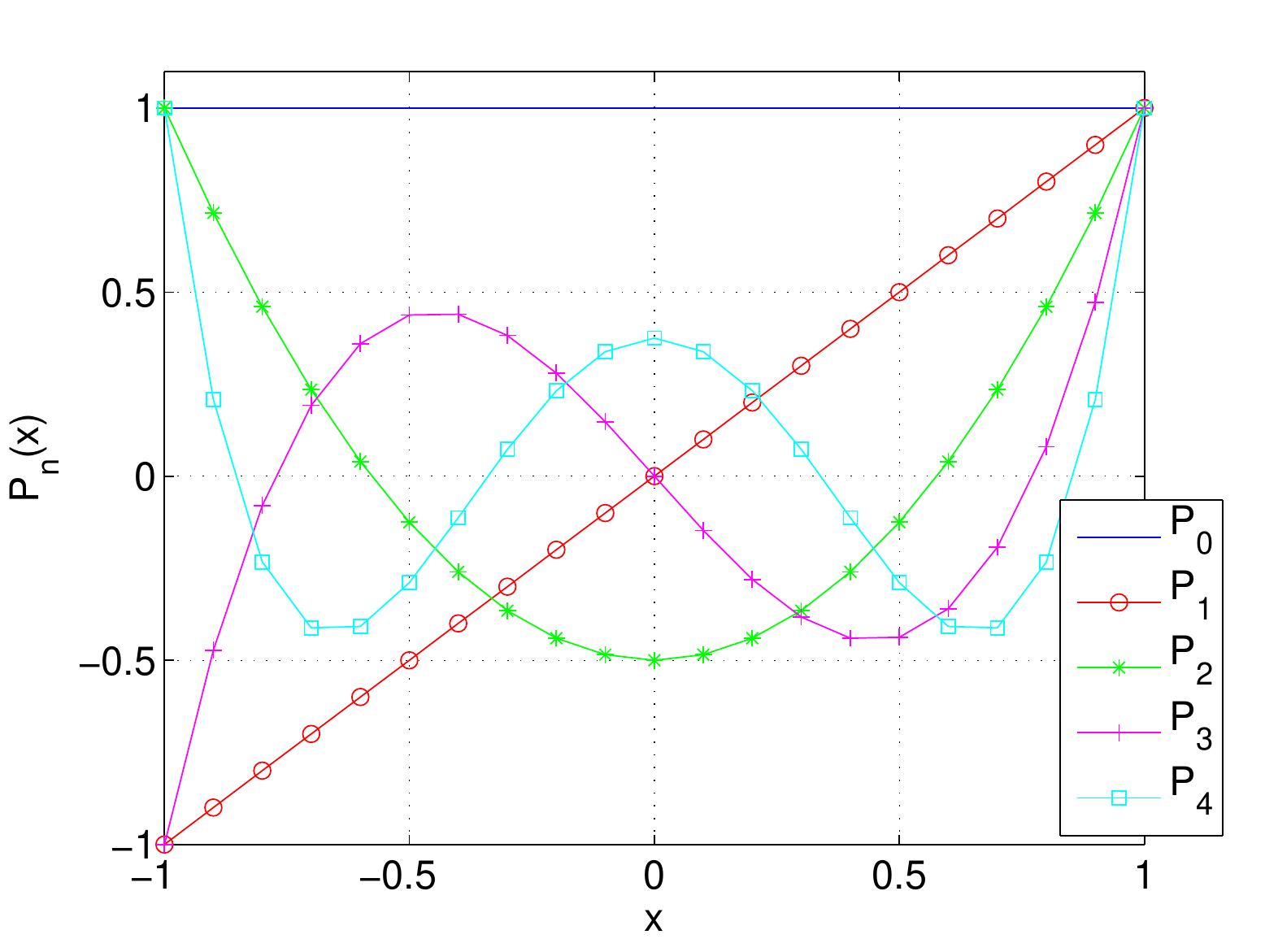} \ \includegraphics[width=65mm]{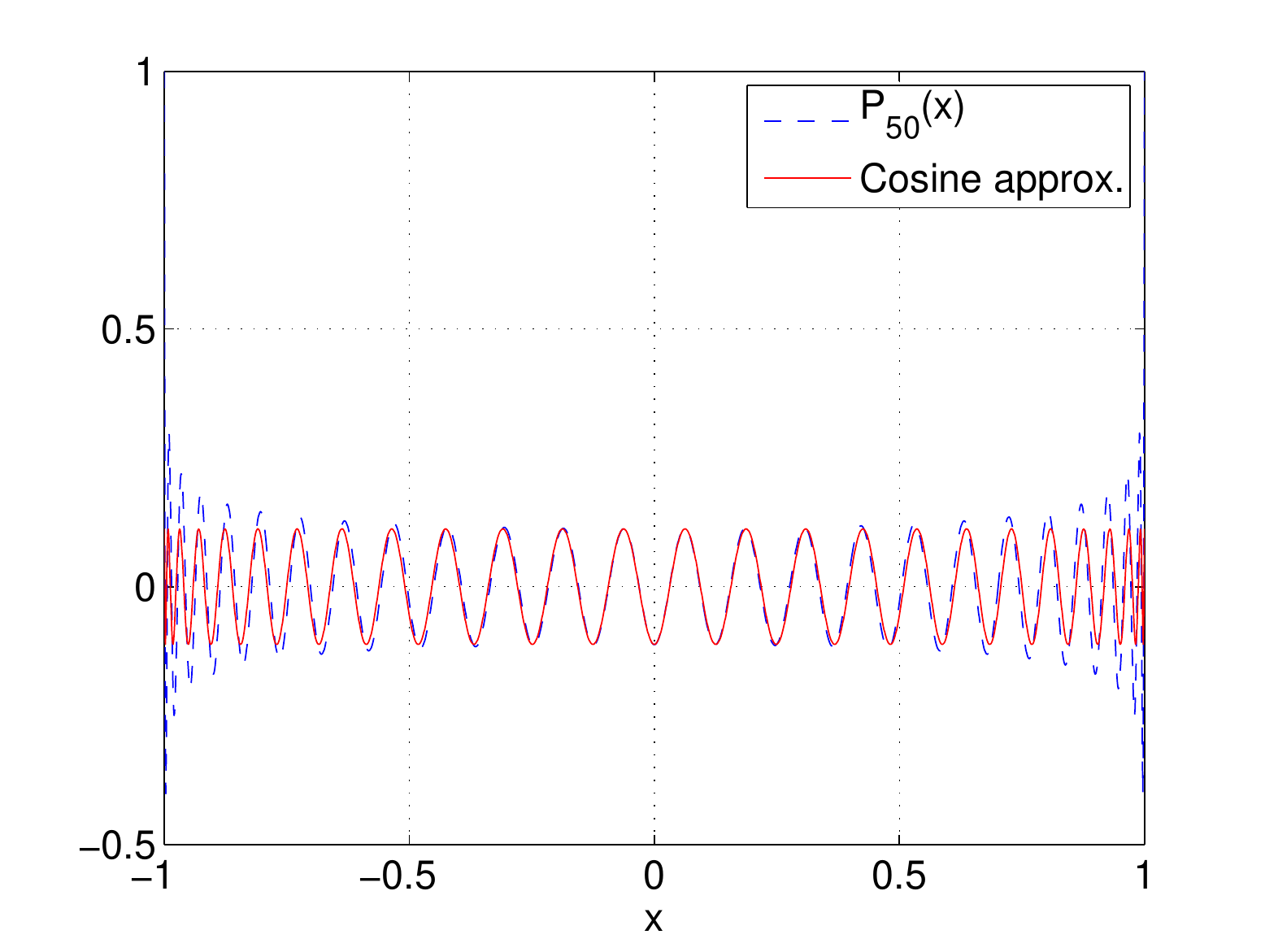} 
\caption{Left: Legendre polynomials $P_n(x)$ for $n=0$ to 4. Right: Legendre polynomial for $n=50$ and its approximation as a cosine function.}
\label{fig:pn}
\end{figure}

The Taylor expansion as powers of $r$ of the generating function $g(x,r)$ gives the following compact expression for the Legendre polynomials, valid for $|x|\le 1$, known as Rodrigues formula
\be
P_n(x)\: = \: \frac{1}{2^n n!}\frac{d^n}{dx^n}(x^2-1)^n
\label{eq:rodrigue}
\ee
The first polynomials are $P_0(x)=1$, $P_1(x)=x$, $P_2(x)=\frac{1}{2} (3 x^2 -1)$. A graph for polynomials $P_0$ to $P_4$ is given in Fig.~\ref{fig:pn}. Note that polynomials corresponding to even (resp. odd) degree $n$ are even (resp. odd), and that $P_n(1)=1 \; \forall n$. The number of distinct roots in the interval $[-1, 1]$ is also $n$. At large $n$ polynoms present oscillations whose frequency and amplitude increase as $|x|\rightarrow 1$. Using eq.~8.10.7 of  Abramowitz \& Stegun (\cite{abramowitz}), it is possible to obtain the following approximation near the origin $x\simeq 0$
\be
P_n(x)\: \simeq \: \frac{n!}{\Gamma(n+1.5)}\: \sqrt{\frac{2}{\pi}} \cos\left[n \left(x-\frac{\pi}2\right)\right]
\label{eq:pncos}
\ee
Fig. \ref{fig:pn} displays the graph of $P_{50}$ and its cosine approximation of Eq.~\ref{eq:pncos}. A large number of properties and relations between Legendre polynomials are found in Abramowitz \& Stegun (\cite{abramowitz}).

\subsection{Legendre differential equation}
From the Rodrigues formula of Eq. \ref{eq:rodrigue}, it comes that the polynoms $P_n(x)$ are solution of the following equation
\be
(1-x^2) y''-2x y'+n(n+1) y=0
\label{eq:legendreDE}
\ee
It is the Legendre DE and it has two independent base solutions: polynoms $P_n(x)$, which are regular on the interval $[-1, 1]$, and functions $Q_n(x)$, infinite at $x=\pm 1$, known as Legendre functions of the second kind. As the equation is linear, any solution is a linear combination of $P_n$ and $Q_n$. However there is no term in $Q_n$ if the solution is to remain finite at $x=\pm 1$. 
\subsection{Fourier-Legendre series}
As for Fourier series, it is possible to make expansion of a function $f$ as a sum of Legendre polynomials. This concerns only functions with bounded support since $P_n(x)$ is defined for $|x|\le 1$. This is generally the case in physics or signal processing.
Following the ideas on scalar products presented in Section~\ref{par:scalarprod}, one can define a scalar product suited to Legendre polynomials (Kaplan~\cite{kaplan}). Let $f$ and $g$ two functions defined on the interval $[-1,1]$, then
\be
\langle f,g\rangle \: = \: \int_{-1}^1 f(x)\, g(x)\, dx
\label{eq:legendrescalarprod}
\ee
Note that this definition of the scalar product is different from the one given in Eq.~\ref{eq:scalarprodfou}, valid for Fourier expansion. It can be demonstrated, using multiple integration by parts, that 
\be
\langle P_n,P_m\rangle = \frac{1}{n+\frac 1 2}\, \delta{mn}
\ee
The family of polynoms ${P_n}$ form a complete orthogonal base of the space of functions square-summable on $[-1,1]$. The base is not orthonormal since the norm $\langle P_n,P_n\rangle$ is not 1. Hence it is possible to expand a function $f$ as the series 
\be
f(x)=\sum_{n=0}^\infty c_n \, P_n(x)
\ee
with 
\be
c_n=\left(n+\frac1 2\right)\; \int_{-1}^1 f(x)\, P_n(x)\, dx
\label{eq:coeflegendre}
\ee
This sum is the Fourier-Legendre expansion of the function $f$. Its form is similar to a Fourier series (Eqs~\ref{eq:seriefourier} and~\ref{eq:coefcn}). As in the case of Fourier series, polynoms of high degree (large $n$ in the Fourier-Legendre decomposition) are associated to short-scale variations of the function $f$. Indeed Legendre polynomials for large $n$ approximate to high frequency cosine functions, as displayed in Fig.~\ref{fig:pn}. 

\subsection{Example of Fourier-Legendre expansion}
\begin{figure}
\includegraphics[width=65mm]{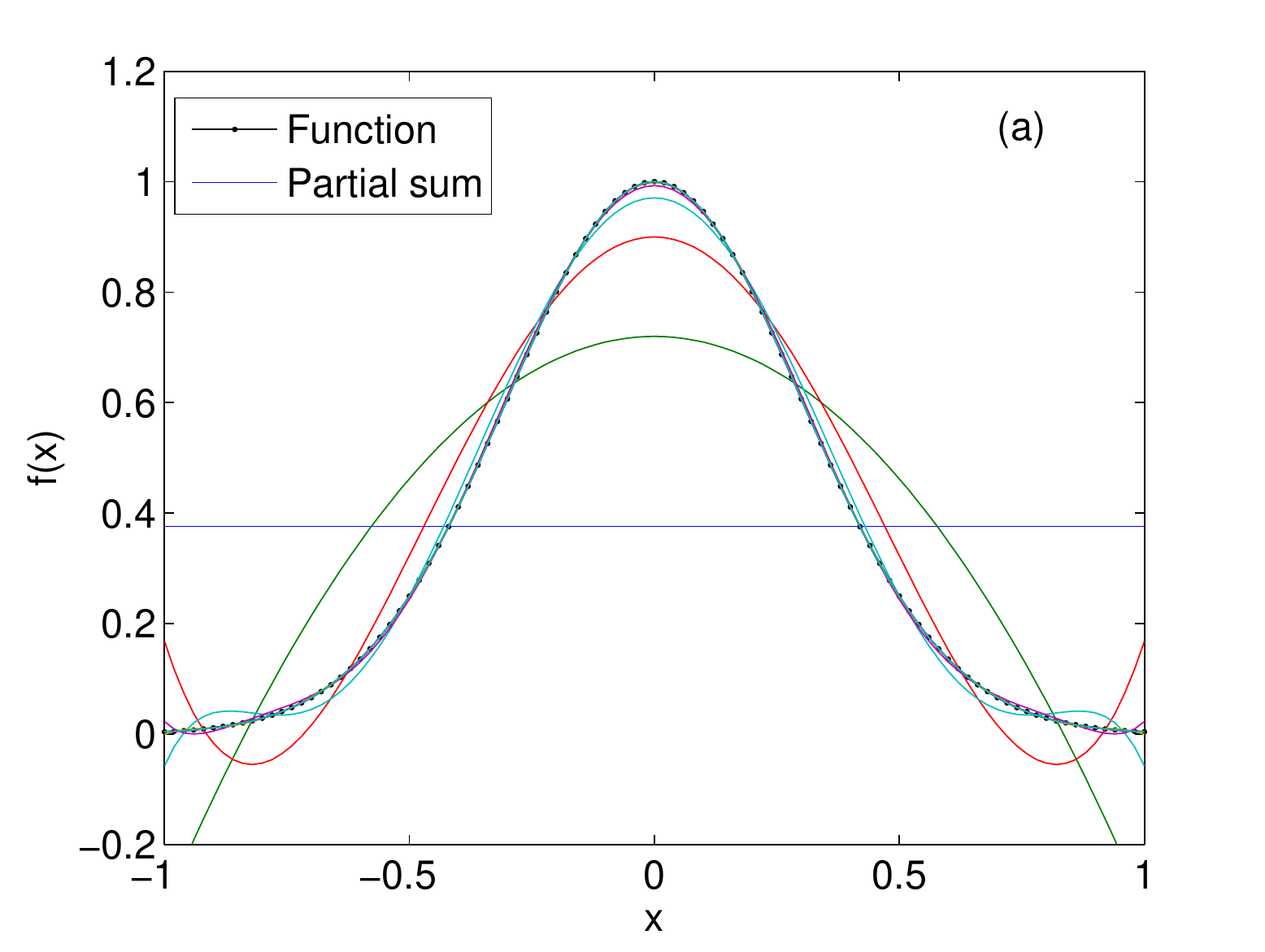}\ \includegraphics[width=65mm]{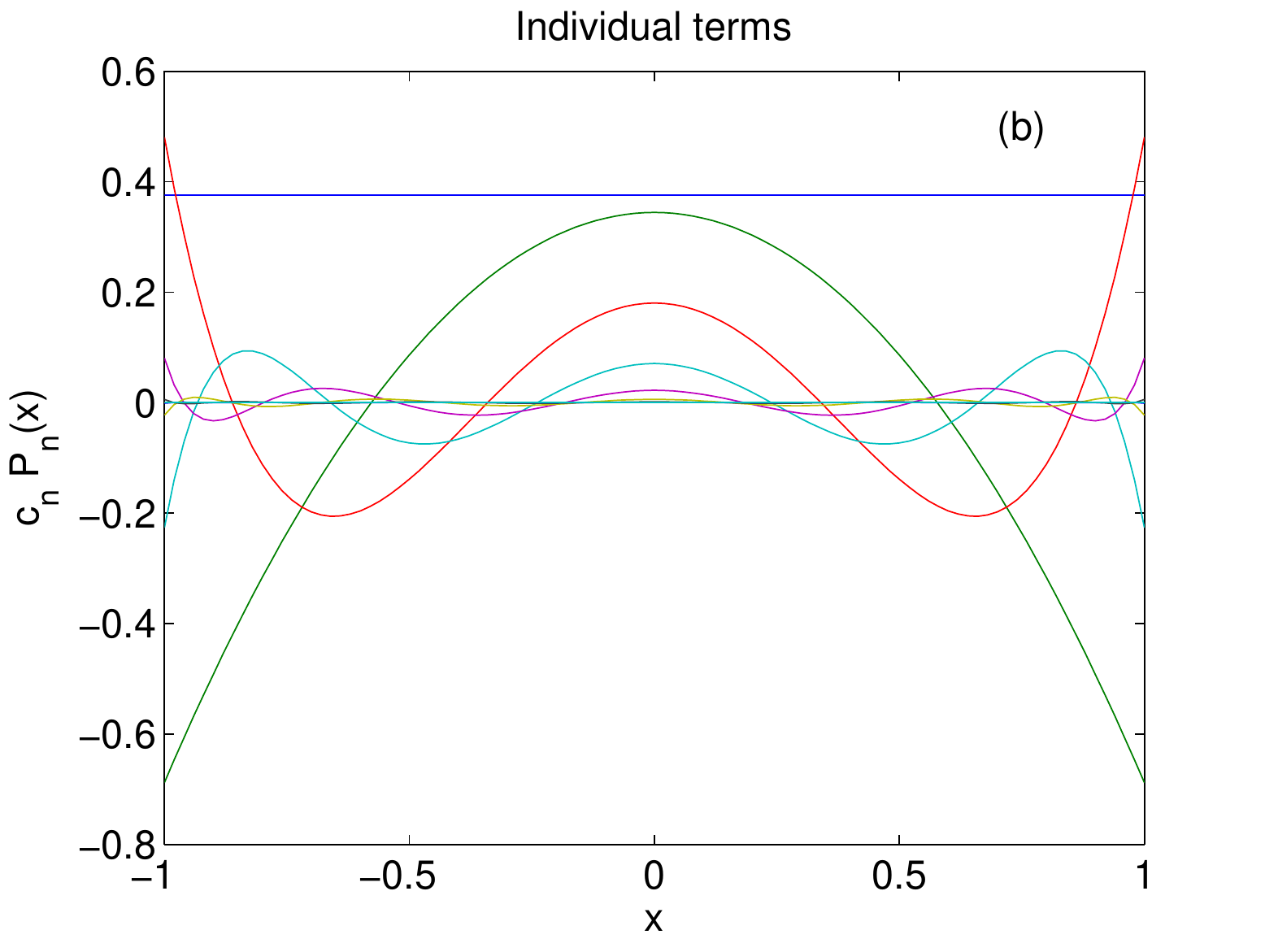}
\includegraphics[width=65mm]{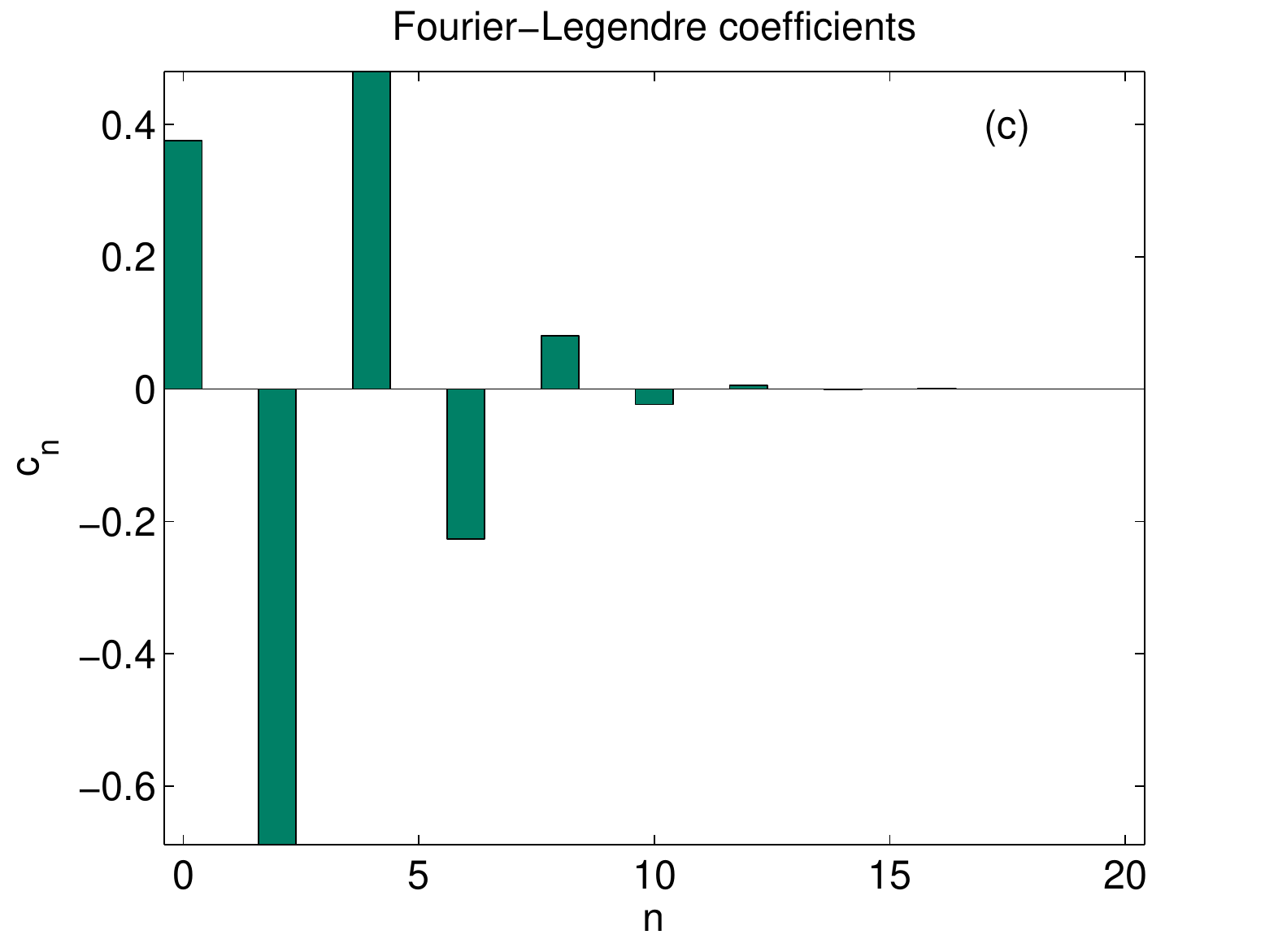}\ \includegraphics[width=65mm]{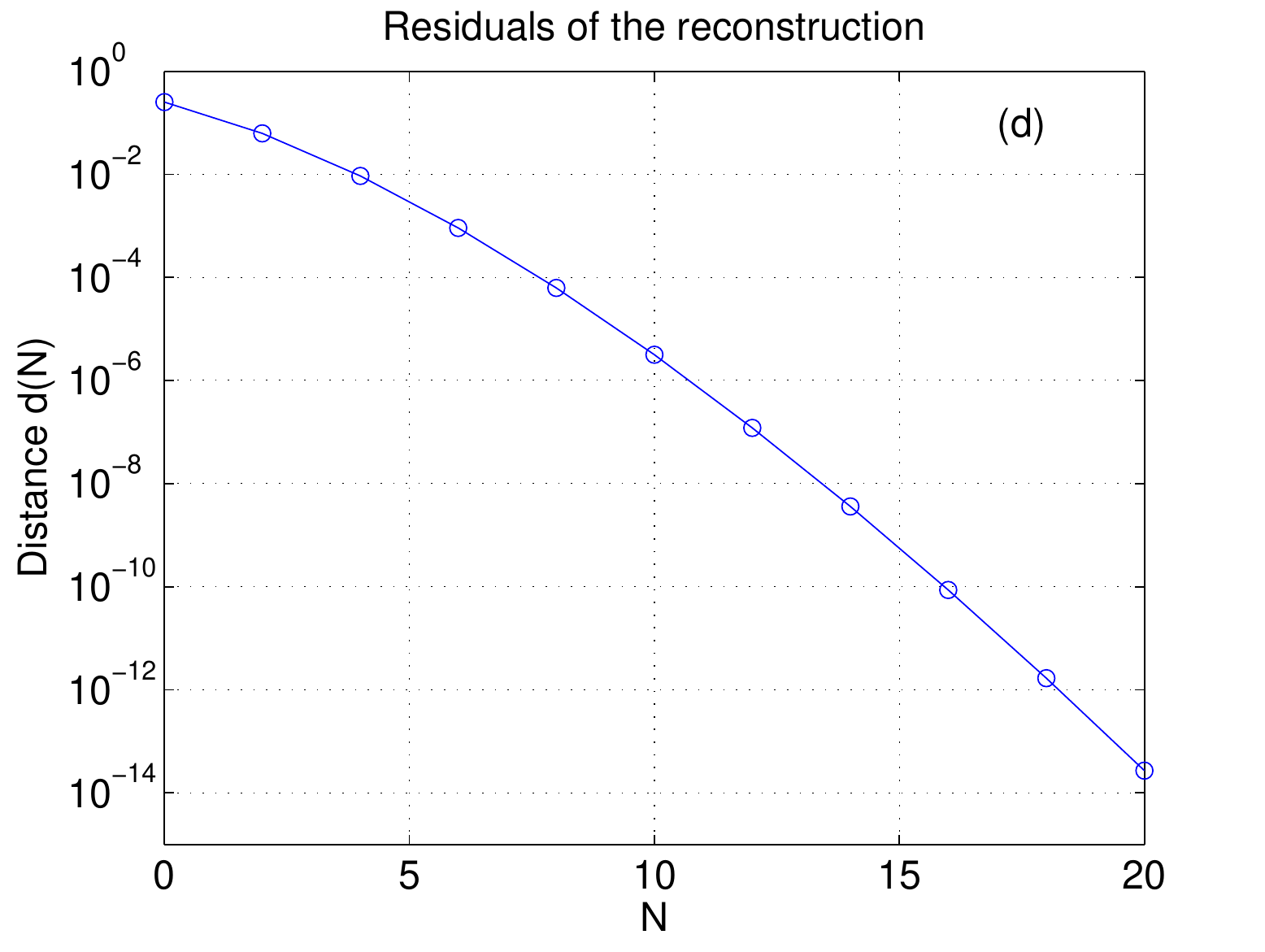}
\caption{Fourier-Legendre reconstruction of the Gaussian function of Eq.~\ref{eq:gaussian} with a$=0.3$. (a) Function and partial sums $S_N(x)$ up to $N=20$. (b) Individual terms of the series  $c_n P_n(x)$ for $n=1$ to 20. (c) Graph of $c_n$ versus $n$ (all odd terms vanish). (d) Euclidian distance $d(N)$ between the function and $S_N$.}
\label{fig:sfl}
\end{figure}

We consider a Gaussian function $f(x)$ having value
\be
f(x)=\exp\left( -\frac{x^2}{2 a^2} \right)
\label{eq:gaussian}
\ee
for $|x|\le 1$ and 0 elsewhere. We took $a=0.3$ for this example. Fourier-Legendre coefficients $c_n$ are 0 for odd $n$. Nonzero coefficients were calculated numerically from Eq.~\ref{eq:coeflegendre}. We then computed partial Fourier-Legendre sums defined as
\be
S_N(x)=\sum_{n=0}^{N} c_n \, P_n(x)
\ee
Fig. \ref{fig:sfl}a shows the graphs of partial sums $S_N(x)$ for $N=0$ to 20. The sum converges nicely towards the function $f$, and the two graphs are coincident for $N\ge 10$ (i.e. a sum of 5 nonzero terms). Fig~\ref{fig:sfl}b displays the individual terms of the series $c_n \, P_n(x)$.  Fig~\ref{fig:sfl}c is a plot of $c_n$ as a function of $n$, it is then the Fourier-Legendre spectrum of $f$ and contains the same information.

The convergence of $S_N$ towards $f$ is shown by Fig. \ref{fig:sfl}d. For each value of $N$, we computed the Euclidian distance between $f$ and $S_N$, defined as
\be
d(N)=\int_{-1}^1 (f(x)-S_N(x))^2\, dx
\ee
The convergence is here very fast, and the sum of only 10 nonzero terms ($N=20$) is enough to obtain a distance of $10^{-14}$ (the starting point was $d(0)\simeq 0.2$). Every term added to $S_N$ divides the distance by about 50.

\subsection{Solution for Laplace equation in spherical coordinates with azimuthal symmetry}
\label{par:laplacesym}
\begin{figure}
\begin{center}
\includegraphics[width=60mm]{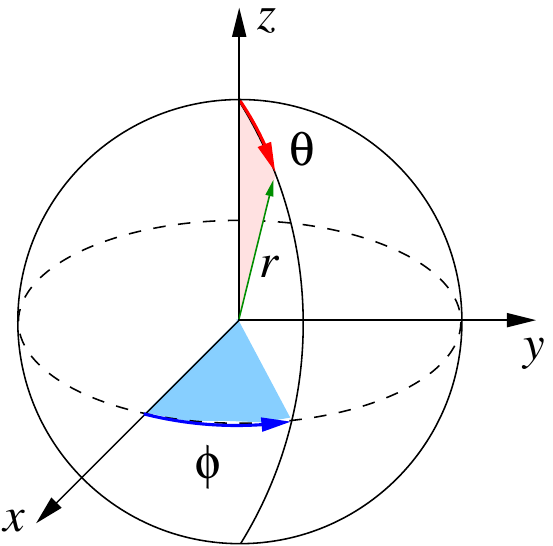} \end{center}
\caption{Spherical coordinate system: $r$ is the distance from the center, $\theta$ the colatitude (zero at North pole) and $\phi$ is the longitude (zero on the $x$ axis).}
\label{fig:shericoord}
\end{figure}

As it was said in Sect.~\ref{par:introlegendre}, Legendre polynomials were introduced to express the Newtonian potential of a mass or charge ditribution. This connection with physics can be made in a more general way through the DE of the potential. We consider here a charge distribution {\em presenting the azimuthal symmetry around the $z$ axis}. And we will express the potential in spherical coordinates $(r, \theta, \phi)$ of the Fig.~\ref{fig:shericoord} with the condition $V$ independent of $\phi$ (azimuthal symmetry).

The potential obeys Laplace's equation
\be
\Delta V =0
\label{eq:laplace}
\ee
which becomes, in spherical coordinates 
\be
\frac 1{r^2}\frac{\partial}{\partial r}\left(r^2 \frac{\partial V}{\partial r}\right)\: + \:
\frac 1{r^2 \sin \theta} \frac{\partial}{\partial \theta}\left(\sin\theta \frac{\partial V}{\partial \theta}\right)\: = 0
\label{eq:laplacespher}
\ee
We apply the technique of separation of variables (see Sect.~\ref{par:fourierserDE}) and seek solutions of the form
\be
V(r,\theta)=f(r).g(\theta)
\ee
Eq.~\ref{eq:laplacespher} can be rewriten as
\be
\frac 1 f \frac d{dr}(r^2 f')\: = \: -\frac 1{g \sin \theta} \frac d{d\theta}(\sin\theta \, g')
\ee
the two sides of this equation must be constant since they depend on different variables, so that
\be
\label{eq:laplacesystemaz}
\left[
\begin{array}{l}
\displaystyle \frac 1 f \frac d{dr}(r^2 f')=\mbox{Cte}=\alpha (\alpha+1)\\ \\
\displaystyle \frac 1{g \sin \theta} \frac d{d\theta}(\sin\theta \, g')=-\alpha (\alpha+1)
\end{array}
\right.
\ee
where the constant was written as $\alpha (\alpha+1)$ for convenience. The solution for $f$ is of the form
\be
f(r)=A_1\, r^\alpha+A_2\, r^{-1-\alpha}
\ee
where $A_1$ and $A_2$ are constants. For $g$, the substitution $x=\cos\theta$ gives
\be
(1-x^2) g''-2 x g'+\alpha(\alpha+1) g=0
\ee
which has regular solutions at $x= 1$ if $\alpha=n$ with $n$ a positive integer. This equation is then exactly the Legendre's DE of Eq.~\ref{eq:legendreDE}. Hence $g$ is the linear combination of the base solutions corresponding to any $n$
\be
g(\theta)=\sum_{n=0}^\infty c_n P_n(\cos\theta)
\ee
and the complete solution for the potential is the Fourier-Legendre series
\be
V(r,\theta)=\sum_{n=0}^\infty (A_n\, r^n+B_n\, r^{-1-n})\: P_n(\cos\theta)
\label{eq:sollaplaceaz}
\ee
Coefficients $A_n$ and $B_n$ are determined by boundary conditions. This example shows how a Fourier-Legendre series appear as the natural expansion of $V$ in this coordinate system with the condition of azimuthal symmetry.

\subsection{Other orthogonal polynomials}
\paragraph{The Sturm-Liouville problem}
\label{par:sturm}

Fourier and Fourier-Legendre series are indeed particular cases of a more general problem explored by Sturm and Liouville in the early 1800 (Liouville,~\cite{liouville}). See the book by Brown~\& Churchill,~(\cite{brown}) for a modern presentation. The Sturm-Liouville (SL) theory explores solutions of second-order DE of the form
\be
\frac{d}{dx}\left[p(x) y'\right]+(\lambda \rho(x)+q(x)) y=0
\ee
where $y(x)$ is the solution, defined on an interval $x\in[a,b]$ with fixed boundary conditions on $y$ and $y'$ at $x=a$ and $x=b$. The key result of the SL theory is that the problem has a solution for particular values $\lambda_n$ of $\lambda$. And that solutions $y_n$ corresponding to each value of $n$ are orthogonal with respect to the weight function $\rho(x)$. Series expansion on Fourier harmonics or Legendre polynomials can be considered as a special case of the SL boundary value problem. As an example, the Legendre equation of Eq.~\ref{eq:legendreDE} is equivalent to the following form
\be
\frac{d}{dx}\left[(1-x^2) y'\right]+n (n+1) y =0
\ee
i.e. a SL problem with $p(x)=(1-x^2)$, $q(x)=0$, $\rho(x)=1$ and $\lambda_n=n (n+1)$.

SL problems occur frequently in physics since a wide class of phenomena are described by a second order PDE. Depending on the coordinate system or circumstances, one may find solutions as series expansion of different kind of functions. Fourier series are likely to occur in cartesian coordinates while Fourier-Legendre series are often met in spherical coordinates for problems having azimuthal symmetry (Jackson,~\cite{jackson}).

Hereafter we give two examples of orthogonal polynom families, related to particular Sturm-Liouville problems. Hermite and Laguerre polynoms have applications in quantum mechanics (Griffiths, \cite{griffiths}). And in Sect.~\ref{par:bessel} we present another example with series of Bessel functions.

\paragraph{Laguerre polynomials $L_n(x)$}
\ 

\begin{tabular}{l|c}
Definition domain & ${\cal D}=[0, +\infty[$ \\  \\ 
 Weight of the scalar product & $\rho(x)=e^{-x}$\\ \\ 
 Orthogonality & 
$\displaystyle
\langle f,g\rangle \: = \: \int_0^\infty e^{-x}\, L_n(x)\, L_m(x)\, dx\: =\: \delta{mn}
$\\ \\ 
Coefficient determination for \\the expansion  $\displaystyle f(x)=\sum_{n=0}^\infty c_n L_n(x)$ &
$\displaystyle
c_n=\langle f,L_n \rangle \: = \: \int_0^\infty e^{-x}\, f(x)\, L_n(x)\, dx
$ \\ \\ 
Generating function (for $|t|<1$) : &
$\displaystyle
\frac{1}{(1-t)}\: \exp\left(-\frac{xt}{1-t} \right)\; = \; \sum_{n=0}^\infty L_n(x)\, t^n
$\\ \\ 
Rodrigues formula & 
$\displaystyle
L_n(x)=\frac{e^x}{n!}\, \frac{d^n}{dx^n}(x^n e^{-x})
$ \\ \\ 
DE & 
$
x y''+(1-x)y'+n y=0
$ 
\end{tabular}


\paragraph{Hermite polynomials $H_n(x)$}
\ 

\begin{tabular}{l|c}
Definition domain & ${\cal D}=]-\infty, +\infty[$ \\  \\ 
 Weight of the scalar product & $\rho(x)=e^{-x^2}$\\ \\ 
 Orthogonality & 
$\begin{array}{ll}
\langle f,g\rangle & = \: \displaystyle \int_{-\infty}^\infty e^{-x^2}\, H_n(x)\, H_m(x)\, dx\: \\ 
 & =\: \sqrt\pi 2^n n!\,\delta{mn}
 \end{array}
 $\\ \\ 
Coefficient determination for \\the expansion  $\displaystyle f(x)=\sum_{n=0}^\infty c_n H_n(x)$ &
$\displaystyle
c_n \: = \: \frac{1}{\sqrt\pi 2^n n!} \int_{-\infty}^\infty e^{-x^2}\, f(x)\, H_n(x)\, dx
$ \\ \\ 
Generating function  : &
$\displaystyle
\exp(2xt-t^2)\; = \; \sum_{n=0}^\infty H_n(x)\, \frac{t^n}{n!}
$\\ \\ 
Rodrigues formula & 
$\displaystyle
H_n(x)=(-1)^n e^{x^2}\, \frac{d^n}{dx^n}(e^{-x^2})
$ \\ \\ 
DE & 
$
y''-2xy' +2n y=0
$ 
\end{tabular}

%
%
\section{Spherical harmonics}
\label{par:SH}
\subsection{Associated Legendre functions}
The associated Legendre functions are defined as
\be
P_l^m(x)=(-1)^m (1-x^2)^{m/2} \frac{d^m}{dx^m} P_l(x)
\label{eq:asslegendre}
\ee
with $m$ a positive integer and $m\le l$. The relation 
\be
P_l^{-m}(x)=(-1)^m \frac{(l-m)!}{(l+m)!} P_l^m(x)
\ee
makes it possible to define $P_l^m$ for $-l\le m \le l$. For $m=0$ we have
\be
P_l^0(x)=P_l(x)
\ee
Legendre associated functions are not polynoms if $m$ is odd. They can be considered as a generalization of Legendre polynomials.  Orthogonality relations exist for the $P_l^m$, for fixed $l$ or $m$ (Abramowitz~\& Stegun,~\cite{abramowitz}).

As for Legendre polynomials, the $P_l^m$ functions obey a Sturm-Liouville DE:
\be
(1-x^2) y''-2x y'+\left[l(l+1)-\frac{m^2}{1-x^2} \right]y=0
\label{eq:assoslegendreDE}
\ee
This equation is the {associated Legendre DE}. It has regular solutions at $x=1$ if $-l\le m \le l$.

\subsection{Laplace equation and spherical harmonics}
In Sect. \ref{par:laplacesym} we showed that the general solution of the Laplace equation in spherical coordinates is a Fourier-Legendre series in case of azimuthal symmetry. In the general case where the potential $V$ is a function of the three coordinates $(r,\theta,\phi)$, the Laplace equation expresses as
\be
\frac 1{r^2}\frac{\partial}{\partial r}\left(r^2 \frac{\partial V}{\partial r}\right)\: + \:
\frac 1{r^2 \sin \theta} \frac{\partial}{\partial \theta}\left(\sin\theta \frac{\partial V}{\partial \theta}\right)\: +\:
\frac 1{r^2 \sin \theta} \frac{\partial^2 V}{\partial \phi^2}
= 0
\label{eq:laplacesphergen}
\ee
Seeking solutions of the form $V(r,\theta,\phi)=f(r).g(\theta).h(\phi)$ gives a system of 3 DE similar to Eq.~\ref{eq:laplacesystemaz}. The equation for $f$ in unchanged. The equation for $h(\phi)$ is an harmonic equation whose solution takes the form
\be
h(\phi)=\alpha_1 e^{i m \phi}+\alpha_2 e^{-i m \phi}
\ee
with $m$ an integer (since the potential must have $2\pi$ periodicity in $\phi$), and $\alpha_1$ and $\alpha_2$ complex coefficients. The DE for $g$ is slightly different from the azimuthal symmetry problem (second equation of the system~\ref{eq:laplacesystemaz})

\be
\frac 1{\sin \theta} \frac d{d\theta}(\sin\theta \, g')+\left(\alpha (\alpha+1) -\frac{m^2}{\sin^2\theta}\right) g =0
\ee
which takes the form of the associated Legendre DE (Eq.~\ref{eq:assoslegendreDE}) with the substitution $x=\cos\theta$. 
Thus, $g$ is an associated Legendre function $P_l^m(\cos\theta)$. And the final solution of the Laplace equation in spherical coordinates takes the form of the following series
\be
V(r,\theta,\phi)=\sum_{l=0}^\infty \sum_{m=-l}^l \left( A_{lm} r^l + B_{lm} r^{-1-l}\right) Y_l^m(\theta,\phi)
\label{eq:sollaplaceylm}
\ee
where
\be
Y_l^m(\theta,\phi) = \: \sqrt{\frac{2 l+1}{4\pi} \frac{(l-m)!}{(l+m)!}}\; P_l^m(\cos\theta)\: e^{i m\phi}
\ee
is the {\em spherical harmonic} (SH) function, first introduced by Laplace (\cite{laplace1782}), though the
  denomination ``spherical harmonic'' was introduced later (see MacRobert\ \& Sneddon, \cite{macrobert} for more  
  details about the history of the SH). The number $l$ is the {\em degree} of the SH, and the number $m$ is its {\em order} 
  (we recall that $-l\le m\le l$)
  
  Sometimes a multiplicative factor $(-1)^m$ is prepended to the definition of the SH. This factor is the Condon-Shortley 
  phase and may rather be included in the definition of the associated Legendre functions (Eq.~\ref{eq:asslegendre}), as
   it is the case here.
  
We shall see in Sect.~\ref{par:orthoSH} that the SH functions are orthogonal. They are indeed a basis for representing 2D 
functions $f(\theta,\phi)$ defined for fixed $r$ over the sphere. They are the spherical analogue of the 1D Fourier 
series and play a very important role in physics.
\label{par:laplaceSH}
\subsection{Some properties and symmetries}
\begin{figure}
\includegraphics[width=1.02\textwidth]{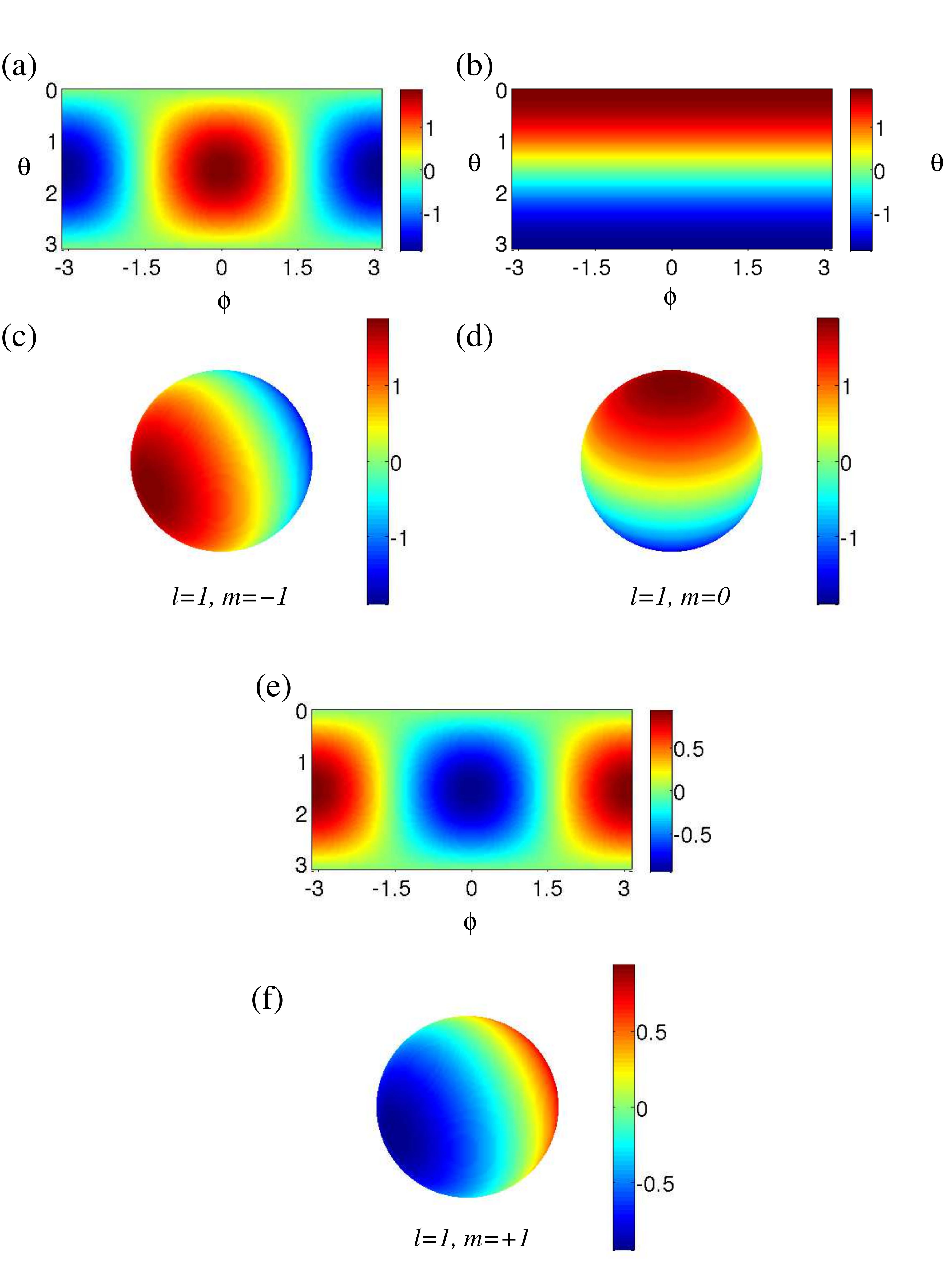}
\caption{Representations of the real part of the SH $Y_l^m(\theta,\phi)$ for $l=1$ and the 3 possible values of $m$. (a) and (c): $m=-1$, (b) and (d): $m=0$, (e) and (f): $m=1$. Graphs (a), (b) and (e) presents false colour plots of the SH in the plane $(\theta,\phi)$. Graphs (c), (d) and (f) displays the same information on the surface of a sphere, as if the sphere was painted with false colours according to the variation of the SH. Angles $\theta$ and $\phi$ are the colatitude and longitude as defined in Fig.~\ref{fig:shericoord}.}
\label{fig:Y1m}
\end{figure}

The first SH are 
\be
\begin{array}{l|l}
\displaystyle Y_0^0=\frac{1}{\sqrt{4\pi}} & 
\displaystyle  Y_2^0=\sqrt{\frac{5}{4\pi}} P_2(\cos\theta) \\ \\
\displaystyle Y_1^0=\sqrt{\frac{3}{4\pi}} \cos\theta & 
\displaystyle Y_2^1 =-\sqrt{\frac{15}{8\pi}} \sin\theta\, \cos\theta\, e^{i\phi} \; = \; -\overline{Y_2^{-1}} \\ \\
\displaystyle Y_1^1=-\sqrt{\frac{3}{8\pi}} \sin\theta\, e^{i\phi} \; = \; -\overline{Y_1^{-1}} & 
\displaystyle Y_2^2 =\sqrt{\frac{15}{32\pi}} \sin^2\theta\, e^{2 i\phi} \; = \; \overline{Y_2^{-2}} 
\end{array}
\ee
SH are functions of the two angles $(\theta,\phi)$ of the spherical coordinates. Various graphic representation are found in the litterature. Fig.~\ref{fig:Y1m} displays the real part of the functions $Y_1^m$ for $l=1$ and $m=-1, 0, +1$. Plots (d), (e) and (f) of  Fig.~\ref{fig:Y1m} are spherical representations: the surface of a sphere is painted in false colours according to the variation of the SH with spherical coordinates angles $(\theta,\phi)$. 

The case $m=0$ corresponds to azimuthal symmetry and the corresponding SH $Y_l^0(\theta)$ is proportionnal to the Legendre polynomial $P_l(\cos\theta)$. The series solution of the Laplace equation in the general case (Eq.~\ref{eq:sollaplaceylm}) is identical to the Fourier-Legendre expansion valid for azimuthal symmetry (Eq.~\ref{eq:sollaplaceaz}).

\begin{figure}
\includegraphics[width=0.75\textwidth]{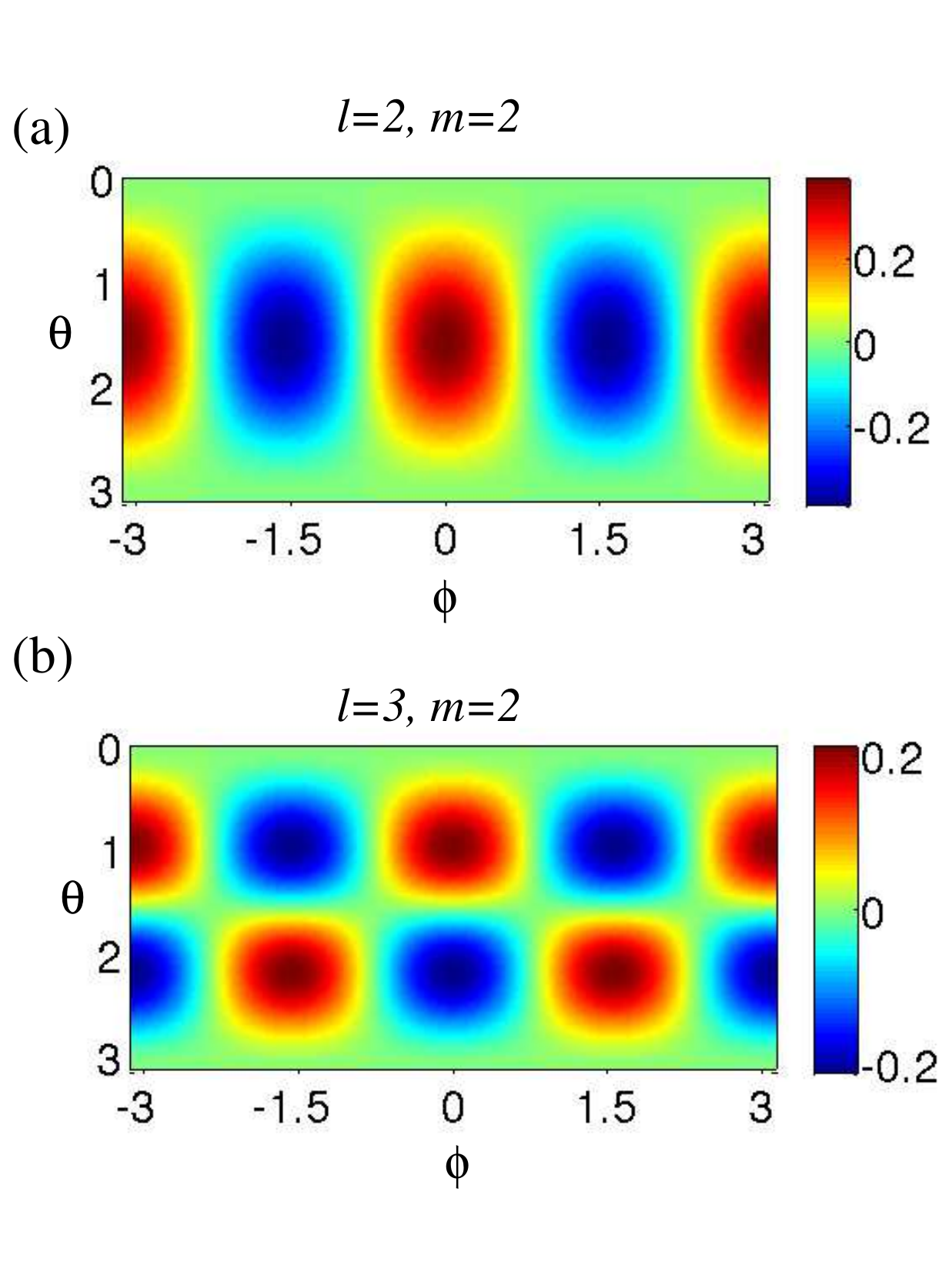}
\caption{False color plots of the real part of the SH $Y_l^m(\theta,\phi)$ for (a): $l=2, m=2$ ($l+m$ even, parity in $\cos\theta$, symmetry plane at the equator) and (b): $l=3, m=2$ ($l+m$ odd, anti-symmetry plane at the equator).}
\label{fig:ylmsym}
\end{figure}

As for Legendre functions, a number of relations exist between SH (Abramowitz \& Stegun,~\cite{abramowitz}). For negative $m$ one can use the expression 
\be
Y_l^{-m}(\theta,\phi)= (-1)^m\, \overline{Y_l^{m}(\theta,\phi)}
\ee
Parity in $\cos\theta$ is the following
\begin{itemize}
\item if $l+m$ is even, the SH is even in $\cos\theta$, i.e. the equator ($z=0$) is a plane of symmetry 
\item if $l+m$ is odd, the SH is odd in $\cos\theta$ and the equator is a plane of anti-symmetry 
\end{itemize}
Examples of both cases are diplayed in Fig.~\ref{fig:ylmsym}. In the SH expansion of the solution to Laplace equation (Eq.~\ref{eq:sollaplaceylm}), a charge distribution presenting a symmetry plane at the equator will create a potential having only even SH in its expansion.

\begin{figure}
\includegraphics[width=0.75\textwidth]{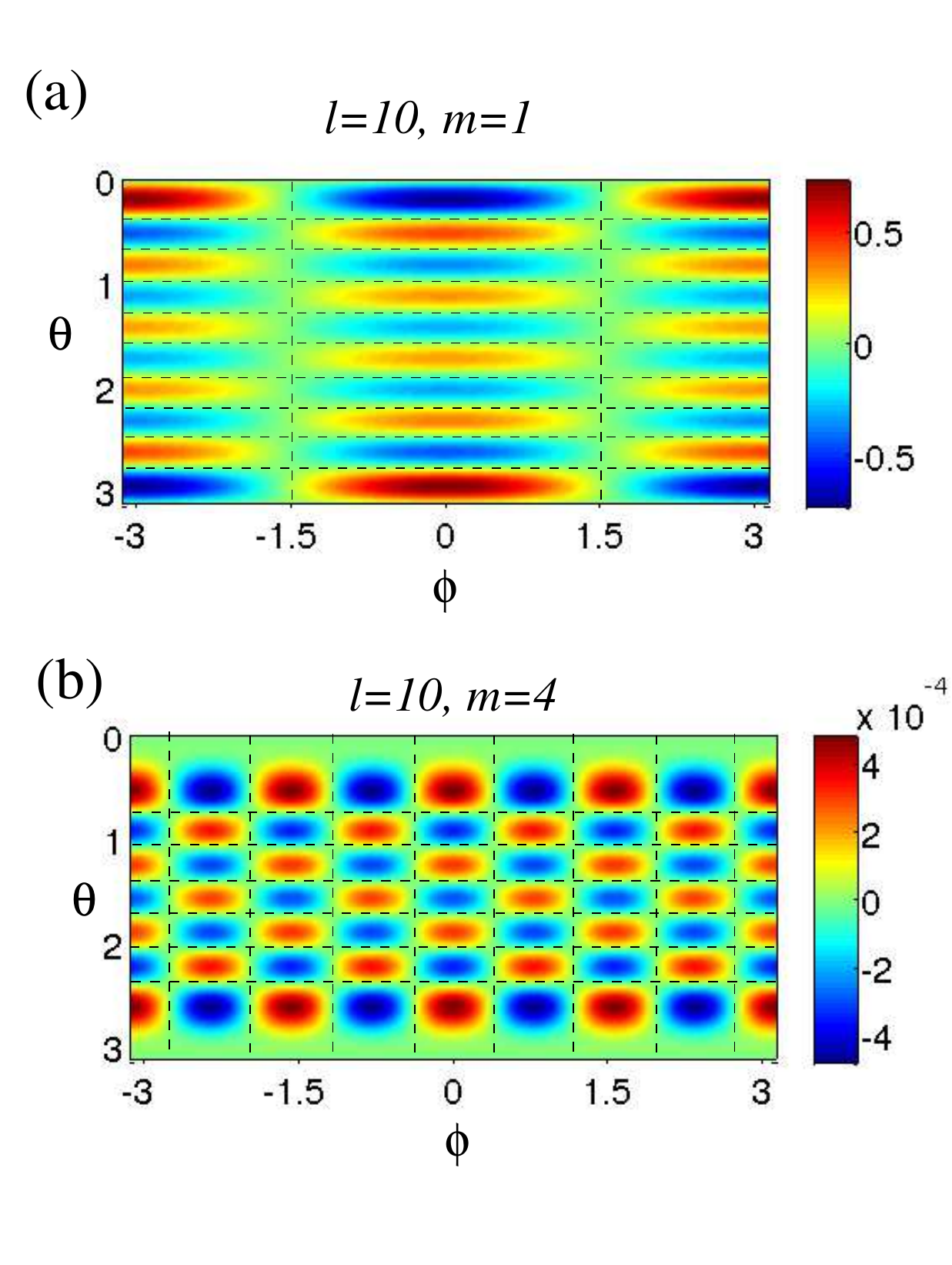}
\caption{False color plots of the real part of the SH $Y_l^m(\theta,\phi)$ for (a): $(l=10, m=1)$ and (b):  $(l=10, m=4)$. Dashed lines are nodal lines corresponding to $\Re [Y_l^m(\theta,\phi)]=0$. The number of nodal lines is $2|m|$ in the $\phi$ direction and $l-|m|$ along the $\theta$ axis.}
\label{fig:ylmnodes}
\end{figure}

The period of the SH $Y_l^m(\theta,\phi)$ in the direction $\phi$ is $\frac{2\pi}{m}$ (if $m\neq 0$). The real part of this function has 2$|m|$ zeros in the interval $\phi\in [0,2\pi[$. In the direction $\theta$, the SH has $l-|m|$ zeros in the interval $\theta\in ]0,\pi[$ (excluding the poles). In the spherical representation, the lines corresponding to $\Re [Y_l^m(\theta,\phi)]=0$ are denoted as {\em nodal lines}. In the direction $\phi$ they are meridian circles passing through the poles. In the direction $\theta$ they are latitude circles parallel to the equator. Fig.~\ref{fig:ylmnodes} shows nodal lines corresponging to the cases $(l,m)=(10,1)$ and $(10,4)$.

SH corresponding to large $|m|$ (resp. large $l-|m|$) have a high angular frequency along the axis $\phi$ (resp. $\theta$). In the SH expansion of a function $f(\theta,\phi)$ they play the same role than high frequency harmonics in a Fourier series: they are associated to short scale variation (small details) of the function $f$.

\subsection{Orthogonality and series expansion}
\label{par:orthoSH}
We consider a function of two angular variables $f(\theta,\phi)$, defined on a sphere in the 3D space, for $\theta\in[0,\pi]$ and $\phi\in[0,2\pi]$. At it was said before, it is possible to make an expansion of $f$ as a sum of SH. One needs to define the following scalar (inner) product between two complex-valued functions $f$ and $g$:
\be
\langle f,g\rangle \: = \: \int_{\theta=0}^\pi\!\!\int_{\phi=0}^{2\pi} f(\theta,\phi)\, \overline{g(\theta,\phi)}\, \sin\theta\, d\theta\, d\phi
\ee
It is easy to show that
\be
\langle Y_l^m, Y_{l'}^{m'}\rangle \: = \: \delta_{l l'}\, \delta_{m m'}
\ee
i.e. that SH forms an orthonormal basis of the space of functions defined on the sphere. The series expansion in SH is then
\be
f(\theta,\phi)\: = \: \sum_{l=0}^\infty \sum_{m=-l}^l a_{lm}\, Y_l^m(\theta,\phi)
\ee
where the coefficients $a_{lm}$ are calculated by the scalar product 
\be
a_{lm}\: = \: \langle f,Y_l^m\rangle \: = \: \int_{0}^\pi\!\!\int_{0}^{2\pi} f(\theta,\phi)\, \overline{Y_l^m(\theta,\phi)}\, \sin\theta\, d\theta\, d\phi
\label{eq:coefylm}
\ee
The norm of the function $f$ is
\be
\langle f,f\rangle \: = \: \sum_{l=0}^\infty \left[\sum_{m=-l}^l |a_{lm}|^2\right] 
\ee
The term between brackets is the usual definition of the {\em angular power spectrum} of $f$.

\subsection{Example}
\begin{figure}
\includegraphics[width=1.02\textwidth]{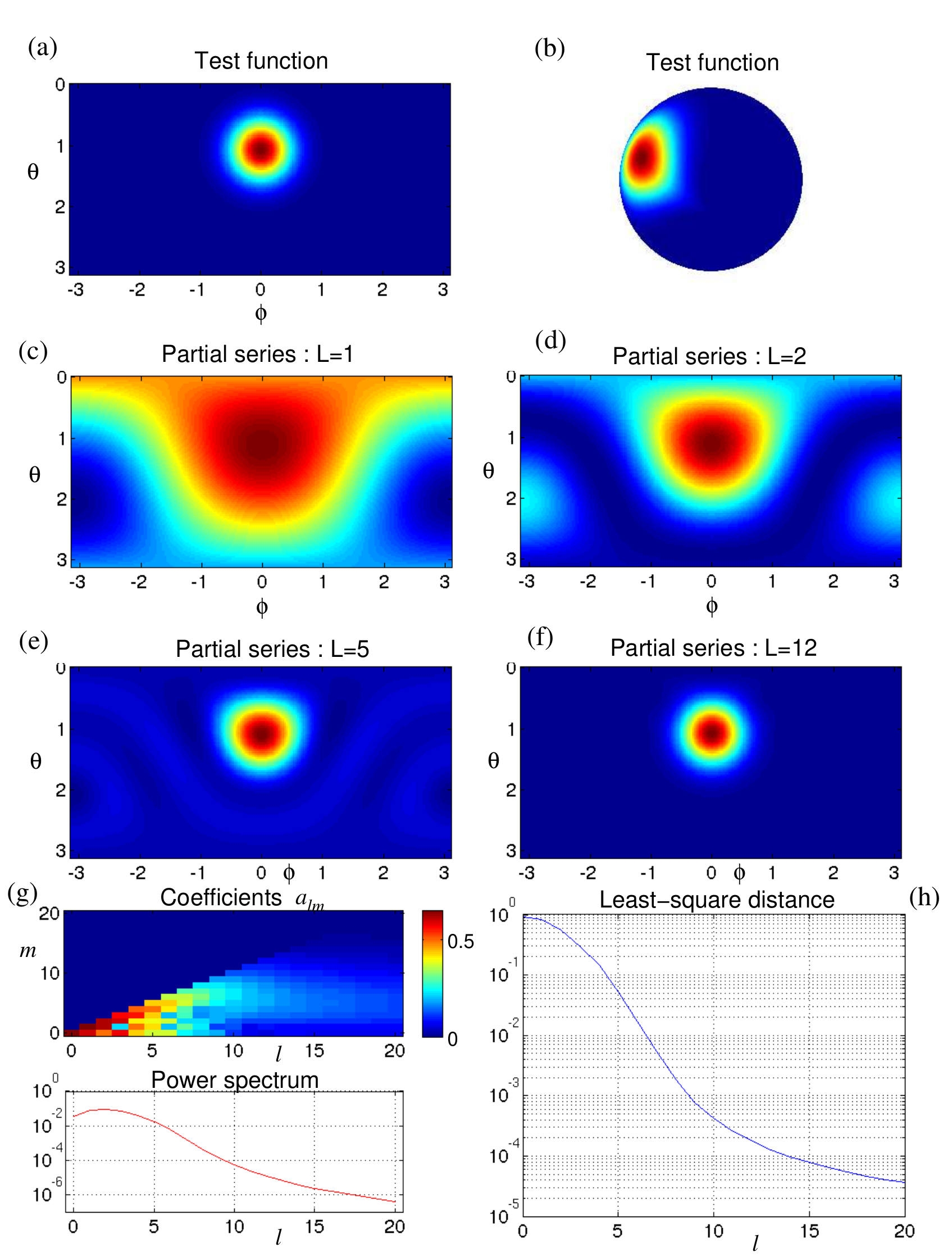}
\caption{Example of SH decomposition of a 2D Gaussian test function (Eq.~\ref{eq:gaussiantestfn}). (a) false color plot of the test function in the plane $(\theta,\phi)$. (b): spherical representation. (c) to (f): plot of the partial SH sum (Eq.~\ref{eq:partialSHsum}) up to $L=1$, 2, 5 and 12. (g): the top graph is a false color plot of the amplitude of the coefficients $a_{lm}$ for $m\ge 0$ in the plane $(l,m)$. The bottom graph is the angular power spectrum as a function of $l$. (h) is the least-square distance between the test function and the partial SH sum.}
\label{fig:ylmexample}
\end{figure}

As an example of SH decomposition, we consider the function
\be
f(\theta,\phi)=\exp\left(-\frac{(\theta-\theta_0)^2+\phi^2}{\delta^2} \right)
\label{eq:gaussiantestfn}
\ee
with $\theta_0=0.5$~rad and $\delta=0.5$~rad. The function is drawn in the $(\theta,\phi)$ plane in Fig.~\ref{fig:ylmexample}a and over a sphere in Fig.~\ref{fig:ylmexample}b. It may represent a bright spot at the surface of a star. Coefficients of the SH decomposition of $f$ were numerically computed using Eq.~\ref{eq:coefylm}. We then computed the truncated SH decomposition defined as
\be
S_L(\theta,\phi)\: = \: \sum_{l=0}^L \sum_{m=-l}^l a_{lm}\, Y_l^m(\theta,\phi)
\label{eq:partialSHsum}
\ee
The results are shown in Fig.~\ref{fig:ylmexample}c, d, e, and f for $L=1$, 2, 5 and 12. For a given $L$ this corresponds to $(L+1)^2$ terms in the sum $S_L$. We can see that the function appears to be well reconstructed for $L=12$. Fig.~\ref{fig:ylmexample}g (top graph) shows the modulus of coefficients $a_{lm}$ in the $(l,m)$ plane, for positive $m$ (negative $m$ verify $a_{l,-m}=(-1)^m\, a_{lm}$ since $f$ in even in $\phi$). Largest coefficients are obtained for small $m$ and $l$, just as it is for 1D Fourier decomposition of a Gaussian signal. The angular power spectrum $\sum_{m=-l}^l |a_{lm}|^2$ is plotted in the bottom graph and confirms this trend.
The convergence of the series is shown in Fig.~\ref{fig:ylmexample}h. It is a plot of the least-square distance between $f$ and $S_L$ as a function of $L$. The distance was normalized to 1 for $L=0$.

SH decomposition is widely used in domains of physics described by a second order PDE such as the Laplace, Schr\"odinger or the wave equation. For example stellar oscillations are often described in terms of standing waves whose angular part in spherical coordinates is a SH (see chapter 8 of Collins,~\cite{collins} for a rewiew about stellar pulsations). Another application is to use SH decomposition for decomposition of functions defined over the sphere. It is indeed an image processing technique analog to 2D Fourier decomposition of an image defined on a rectangle. For example in cosmology, the brightness distribution over the whole sky of the Cosmic Microwave Background (CMB) is analyzed in terms of SH series. The angular power spectrum provides informations about the statistical properties of the CMB. Hinshaw et al.,~ (\cite{hinshaw}) present an extensive analysis of data from the WMAP spacecraft.
%
%
\section{Bessel functions}
\label{par:bessel}
Bessel functions were introduced after the work by Bessel (\cite{bessel}) about the motion of the planets around the Sun. Bessel expressed the position $r(t)$ of the planet as a temporal Fourier series whose coefficients were defined by integrals. In his memoir, Bessel made systematics investigation of the properties of these integrals who now bear his name. An extensive treatise about Bessel functions is that of Watson~(\cite{watson}) and includes an historical introduction (one can also refer to Dukta,~\cite{dukta}).
\subsection{Bessel differential equation}
\label{par:besselDE}
\begin{figure}
\includegraphics[width=65mm]{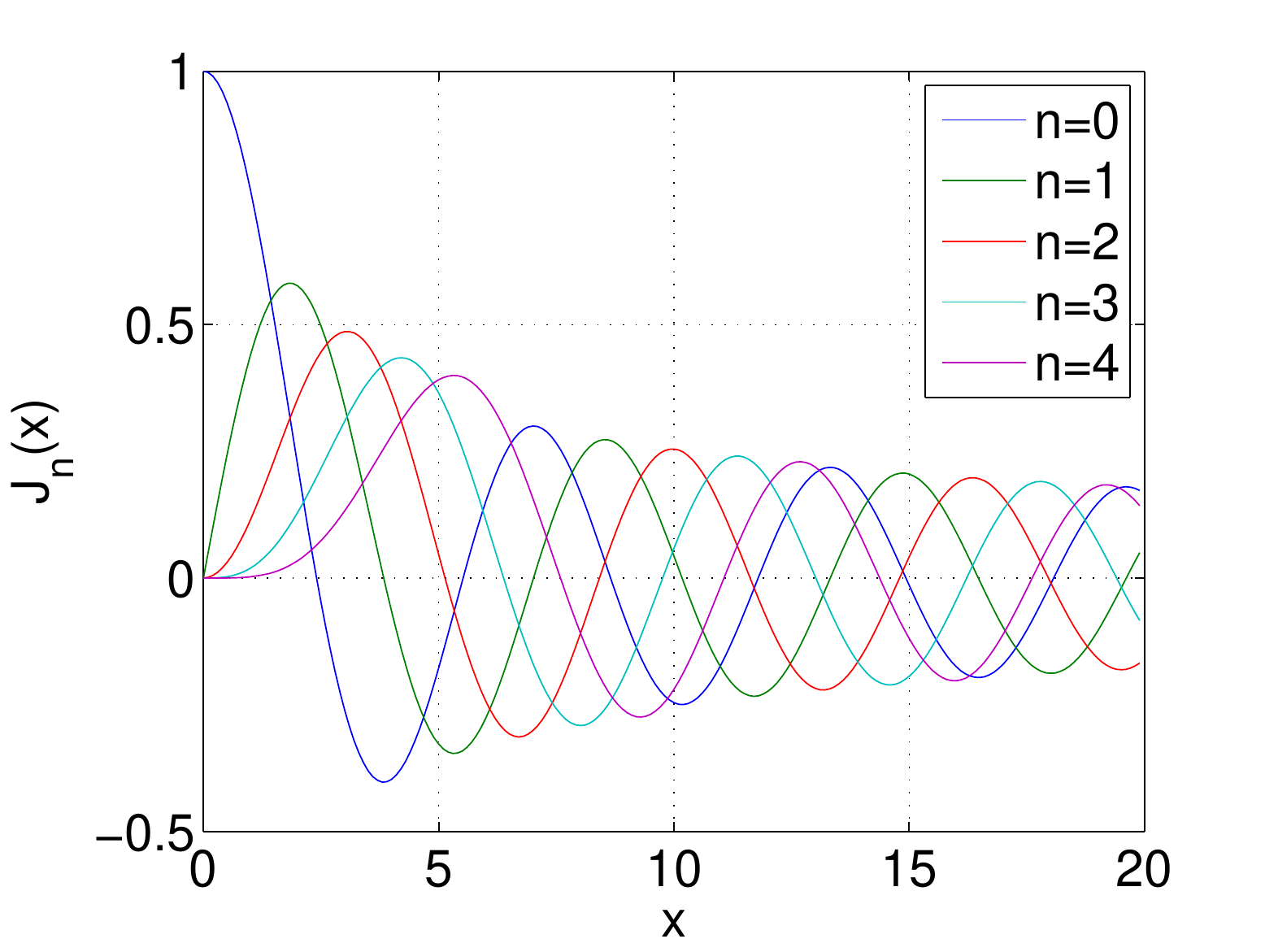}\ \includegraphics[width=65mm]{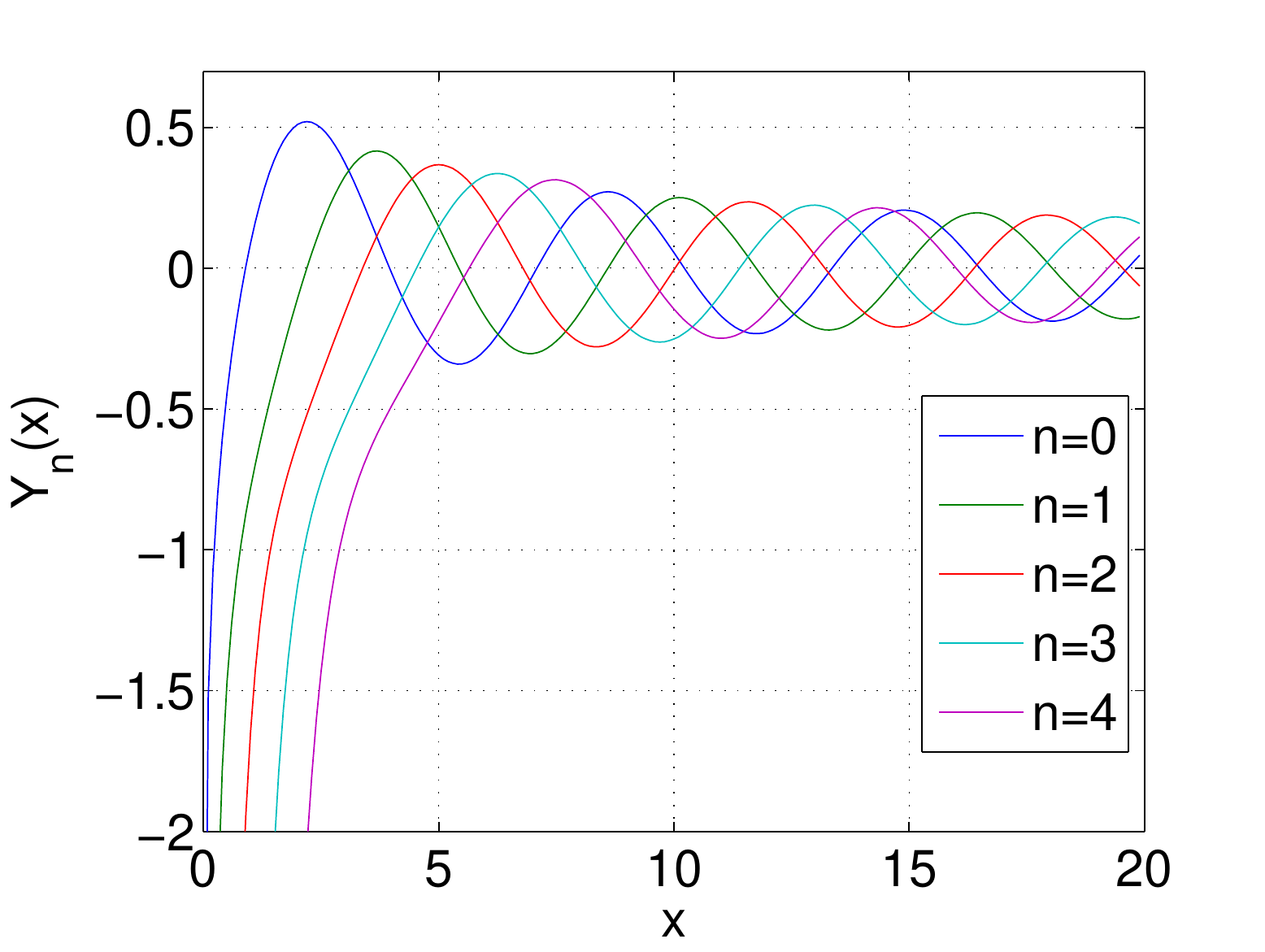}
\caption{Plots of the first Bessel functions for positive argument $x$. Left: $J_n(x)$ for $n=0$ to 4. Right: $Y_n(x)$ for $n=0$ to 4.}
\label{fig:JnYn}
\end{figure}

Bessel functions are often introduced via the Bessel DE
\be
x^2 y''+ x y'+ (x^2-n^2) y=0
\label{besselDE}
\ee
where $n$ is an arbitrary complex number, but in the present paper we shall restrict to integer $n$. The Bessel equation is indeed a Sturm-Liouville problem (see Sect.~\ref{par:sturm}) with $p(x)=x$,  $q(x)=x$, $\rho(x)=-\frac 1 x$ and $\lambda=n^2$. Series solutions of the DE can be obtained by the Frobenius method (Mathews~\& Walker,~\cite{walker}). This equation has two independent base solutions: first kind and second kind  Bessel functions $J_n(x)$ and $Y_n(x)$. The function $J_n$ is regular at $x=0$ for all $n$, while $Y_n$ diverges at $x=0$ and is complex-valued for $x<0$. Plots of the first $J_n$ and $Y_n$ are shown in Fig.~\ref{fig:JnYn}.

The zoology includes also modified Bessel functions $I_n$ and $K_n$, Hankel functions $H_n$ and cylindrical Bessel functions $j_n$ and $y_n$. These functions are of common use in various domains (for example the spherical Bessel functions have applications in quantum mechanics, see chapter 4 of Griffiths~\cite{griffiths}). In this paper we shall focus our presentation to the $J_n$ family.

\subsection{Bessel functions of the first kind  $J_n(x)$}
In Sect.~\ref{par:introlegendre} we introduced the Legendre polynomials by means of a Taylor expansion of a generating function. The same can be made for $J_n$, the generating function $g(x,t)$ being (Watson,~\cite{watson}):
\be
g(x,t)\: =\: \exp\left[\frac x 2 \left(t-\frac 1 t \right)  \right] \: = \: \sum_{n=-\infty}^\infty t^n\, J_n(x)
\label{eq:genbessel}
\ee
Or the equivalent trigonometric form (variable change $t=e^{i\phi}$)
\be
e^{i x \sin\phi}\: = \: \sum_{n=-\infty}^\infty J_n(x)\, e^{i n \phi}
\label{eq:genbesseltrig}
\ee
which is the Fourier series of the function $e^{i x \sin\phi}$, 2$\pi$-periodic in $\phi$, and $J_n(x)$ is the Fourier coefficient calculated by Eq.~\ref{eq:coefcn}
\be
J_n(x)\: = \: \frac{1}{2\pi} \int_0^{2\pi} e^{i x \sin\phi}\, e^{-i n \phi}\, d\phi
\ee
This important formula is an integral representation of $J_n$ which has many applications, for example in optics to calculate  diffraction patterns of screens having rotational invariance (Born \& Wolf, \cite{bornwolf}). It is also very close to the coefficients introduced in the memoir of Bessel (\cite{bessel}).

Basic properties of the functions $J_n$ include
\begin{itemize}
\item Parity: same as $n$ 
\item Negative order: $J_{-n}(x)=(-1)^n J_n(x)$
\item Value at the origin: $J_n(0)=0$ for $n>0$ and $J_0(0)=1$
\item Behaviour for small $x$: $\displaystyle J_n(x) \simeq \frac{1}{n!} \left(\frac x 2  \right)^n$
\item Asymptotic limit ($x\rightarrow \infty$): $\displaystyle J_n(x) \simeq \sqrt{\frac{2}{\pi x}} \cos \left(x-\frac{n \pi} 2 - \frac \pi 4  \right)$
\end{itemize}

All $J_n$ functions behave like damped sinusoids with an infinite number of roots on the real axis. The position of these zeros is not periodic. They play a role in Fourier-Bessel expansion, as it will be further discussed in Sect.~\ref{par:fourierbessel}.

Various expansions of functions in series of Bessel functions exist, see Watson~(\cite{watson}) for a review. Eqs.~\ref{eq:genbessel} and \ref{eq:genbesseltrig} are two examples of Neumann series of the type
\be
f(x)=\sum_{n=-\infty}^\infty a_n J_n(x)
\ee
In next section, we focus on a special case of orthogonal expansion involving Bessel functions, which has many connections with physics: the Fourier-Bessel series.
\subsection{Fourier-Bessel series}
\label{par:fourierbessel}
\subsubsection{Vibrations of a drum membrane}
In Sect.~\ref{par:fourierserDE}, \ref{par:laplacesym} and \ref{par:laplaceSH} we showed that a DE representing a particular problem would led to series solutions depending on the geometry: Fourier expansion for rectangular coordinates, Fourier-Legendre or SH series for spherical coordinates. Here we explore solutions of the equation describing oscillations of a drum circular membrane of radius $R$ in polar coordinates $(\rho, \phi)$ (origin at the center of the membrane). The function $s(\rho,\phi,t)$ describing the displacement of the membrane obeys the wave equation
\be
\Delta s-\frac{1}{c^2}\frac{\partial^2 s}{\partial t^2} =  0
\ee
where $c$ is the speed of propagation. We apply the technique of separation of variable (as in Sect.\ref{par:fourierserDE} and \ref{par:laplaceSH}) and look for solutions of the form $s(\rho, \phi, t)=f(\rho).g(\phi).h(t)$. The wave equation is then equivalent to the following system
\be
\left[
\begin{array}{l}
\displaystyle h'' + k^2c^2\, h = 0\\ \\
\displaystyle g'' + n^2 g = 0\\ \\
\displaystyle \rho^2 f'' + \rho f' + (k^2 \rho^2 - n^2) f =0
\end{array}
\right.
\ee
where $k$ and $n$ are constant ($k$ is positive to have oscillatory solutions for $h(t)$, $n$ is integer to ensure 2$\pi$ 
periodicity in $\phi$). A more detailed step-by-step calculation can be found in Asmar~(\cite{asmar}). It can be noticed that the equation for $f$ is indeed a Bessel DE (Eq.~\ref{besselDE}) whose base solution is 
\be
f(\rho)=J_n(k\rho)
\ee
An important condition is the boundary value, i.e. the membrane must be motionless at its extremities ($\rho=R$), so that $f(R)=0$ whatever $n$. Hence $k$ must have the form $k=\frac{\alpha_{nm}}{R}$ with $\alpha_{nm}$ the $m$-th positive root of the function $J_n$. The general solution is the linear combination of base solutions (i.e. {\em modes}) for every possible $m$ and $n$:
\be
s(\rho, \phi, t)\; = \; \sum_{n=-\infty}^\infty \sum_{m=1}^\infty A_{mn}\, J_n\left(\frac{\alpha_{nm} \rho}{R} \right)\, e^{in\phi} \, e^{i\omega_{mn} t}
\ee
with $A_{nm}$ a constant and  $\omega_{mn}=kc=\frac{\alpha_{nm} c}{R}$ the temporal pulsation of the mode. This double sum is indeed a Fourier series for the variable $\phi$ and a Fourier-Bessel series for the variable $\rho$. 
\subsubsection{Fourier-Bessel expansion}
\begin{figure}
\begin{center}
\includegraphics[width=85mm]{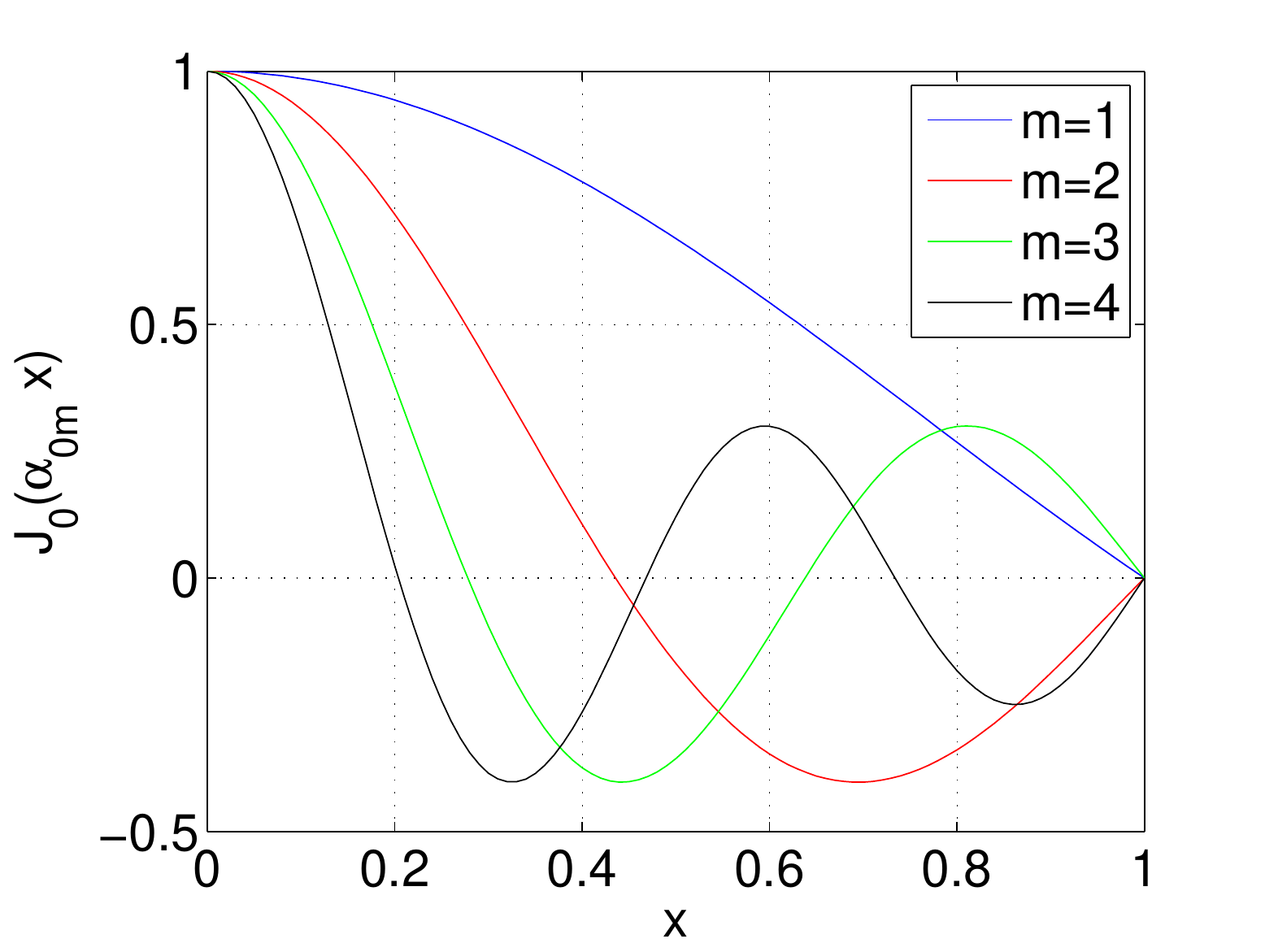}
\end{center}
\caption{Graph of the functions $J_0(\alpha_{0m} x)$ for $m=1$ to 4. Corresponding roots of $J_0$ are $\alpha_{01}=2.4$, $\alpha_{02}=5.5$, $\alpha_{03}=8.7$, $\alpha_{04}=11.8$. These functions are orthogonal with respect to the scalar product defined by Eq.~\ref{eq:pdtscalbessel}. All functions vanish at $x=1$.}
\label{fig:j0alphamx}
\end{figure}

As for Fourier or Fourier-Legendre series, it is possible to make orthogonal expansions of functions as series of Bessel functions. This concerns continuous functions with bounded support $[0, 1]$ who verify $f(1)=0$. One needs to define the following scalar product:
\be
\langle f,g\rangle=\int_0^1 x\, f(x)\, g(x)\, dx
\label{eq:pdtscalbessel}
\ee
Note that this scalar product, suited for Fourier-Bessel expansion, is different from those defined in Eqs~\ref{eq:scalarprodfou} and~\ref{eq:legendrescalarprod}. It can be shown (Watson, \cite{watson}) that
\be
\langle J_n(\alpha_{nm} x), J_n(\alpha_{np} x)\rangle\; = \; \frac{J_{n+1}^2(\alpha_{nm})}{2}\, \delta_{mp}
\ee
so that the functions $j_{m}(x)=J_n(\alpha_{nm} x)$ form a complete set of orthogonal functions. As an example Fig.~\ref{fig:j0alphamx} shows the graph of the base functions $J_0(\alpha_{0m} x)$ for $m=1$ to 4. All of them are stretched version of $J_0$ with a stretching factor depending on the root $\alpha_{0m}$. Any function $f(x)$ continuous on $[0,1]$ with $f(1)=0$ can be expanded as a series of $j_m$
\be
f(x)=\sum_{m=1}^\infty c_m J_n(\alpha_{nm} x)
\ee
with the coefficient
\be
c_m=\frac 2{J_{n+1}^2(\alpha_{nm})} \langle f, j_m\rangle\: = \: \frac 2{J_{n+1}^2(\alpha_{nm})}\: \int_0^1 x\, f(x)\, J_n(\alpha_{nm} x) \, dx
\label{eq:coef_fb}
\ee
This expansion is denoted as Fourier-Bessel expansion of $f$. Note that the choice of $n$, the order of the Bessel function, is arbitrary (or suggested by physics): an infinity of Fourier-Bessel expansions exist for a given function. 

\subsubsection{Exemple of Fourier-Bessel expansion}
We consider the following function for $x\in[0,1]$
\be
f(x)=e^{-3x}\, \cos\left(\frac{3\pi x}{2}  \right)
\label{eq:ftest_fb}
\ee
\begin{figure}
\begin{center}
\includegraphics[width=1.02\textwidth]{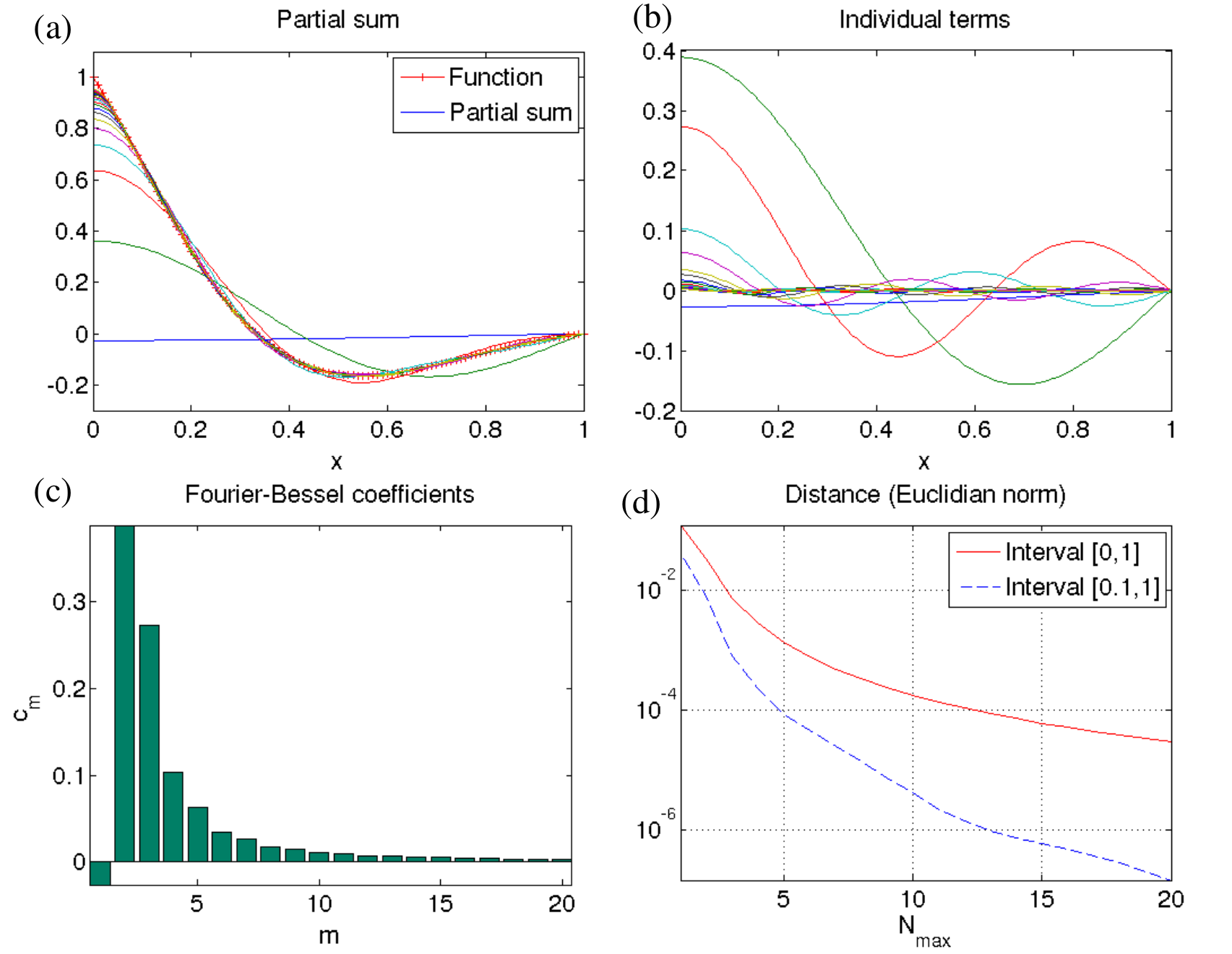}
\end{center}
\caption{Fourier-Bessel reconstruction of the function defined by Eq.~\ref{eq:ftest_fb}, on the base of orthogonal function $J_0(\alpha_{0m} x)$. (a) function and partial sums $S_N(x)$ for $N=1$ to 20. (b) individual terms of the series. (c) graph of $c_m$ versus $m$. (d) Least square distance $d_N$ between the function and the sum for two intervals (Eq.~\ref{eq:dist_fb}).}
\label{fig:fb_ex}
\end{figure}

This function vanishes at $x=1$, as required to make a Fourier-Bessel expansion. Also since $f(0)=1$ we choose to expand it on the basis of $J_0(\alpha_{0m} x)$. Coefficients of the series were calculated numerically from Eq.~\ref{eq:coef_fb}. We computed the partial series
\be
S_N(x)=\sum_{m=1}^N c_m J_n(\alpha_{nm} x)
\ee
up to $N=20$. Results are shown in Fig.~\ref{fig:fb_ex}. The convergence is fairly good for $N=20$, excepted near $x=0$ where more terms are required to fit the function. This is due to the fact that $J_0$ derivative is zero at $x=0$ while it is not for the function. In Fig.~\ref{fig:fb_ex}d, we display the least-square distance $d_N$ between the function and the partial sum as a function of $N$:
\be
d_N=\int_a^1 (f(x)-S_N(x))^2\, dx
\label{eq:dist_fb}
\ee
Two curves are shown: one for $a=0$ (whole interval), the other one for $a=0.1$. As we see, the second curve is two decades below the first for $N=20$, it means that the interval $[0, 0.1]$ contains 99\% of the reconstruction residuals.

Fourier-Bessel expansion can be made using higher order Bessel functions $J_n$, but for this particular example the convergence is very slow near $x=0$ and hundreds of terms are required in the series.

%
%
\section{Conclusion}
This presentation aimed at introducing the mathematical concept of representing a function or a signal as a series of orthogonal functions. We focused on four particular cases: the Fourier series, the Fourier-Legendre series, the Spherical Harmonic decomposition and the Fourier-Bessel expansion. A basic presentation of each set of orthogonal functions  was given, together with a list of references for further reading. The central concept of inner/scalar product of functions was introduced by analogy with ``classical'' vectors in the three-dimensional space. All series expansions were illustrated by a numerical example.

The connexion with physics was enlighted each time, and we showed that the link is often made via differential equations of the physical phenomenom. The geometry often gives a ``natural'' basis for series expansion. Hence Fourier series are often used in rectangular coordinates, SH are the natural base in spherical coordinates and Fourier-Bessel expansions are met in cylindrical or polar coordinates.

A wide variety of series expansions on orthogonal set of functions exist in the field of functionnal analysis, and this presentation is far from exhaustive. For example we did not explore the decomposition of a two-dimensional function in Zernike modes, widely used in adaptive optics (Noll, \cite{noll}).

Also, it is not the sake of this paper to present FFT-like fast algorithms to compute various transforms such Fourier-Legendre or Spherical Harmonics. There is indeed an abundant litterature on this topic and the reader may refer to O'Neil et al. (\cite{oneil}) and references therein.

This paper could hopefully be used as a starting point or a reminder, in particular for students. The reference list gives a number of classical textbooks with much details and exercices.


\end{document}